\documentclass[fleqn,usenatbib]{mnras}

\usepackage{newtxtext,newtxmath,comment}
\usepackage[normalem]{ulem}

\usepackage[T1]{fontenc}
\usepackage{ae,aecompl}

\usepackage{graphicx}	
\usepackage{amsmath}	

\title[SIDM and Baryons in MW Halos]{Unraveling the interplay between SIDM and baryons in MW halos: defining where baryons dictate heat transfer}

\author[J. C. Rose et al.]{
Jonah C. Rose$^{1}$\thanks{E-mail: j.rose@ufl.edu},
Paul Torrey$^{1}$,
Mark Vogelsberger$^{2}$,
Stephanie O'Neil$^{2}$
\\
$^{1}$Department of Astronomy, University of Florida, Gainesville, FL 32611, USA \\
$^{2}$Kavli Institute for Astrophysics and Space Research, Massachusetts Institute of Technology, 70 Vassar St., Cambridge, MA 02139, USA
}

\date{Accepted XXX. Received YYY; in original form ZZZ}

\pubyear{2022}

\begin{document}
\label{firstpage}
\pagerange{\pageref{firstpage}--\pageref{lastpage}}
\maketitle

\begin{abstract}
We present a new set of cosmological zoom-in simulations of a MW-like galaxy which for the first time include elastic velocity-dependent self interacting dark matter (SIDM) and IllustrisTNG physics.
With these simulations we investigate the interaction between SIDM and baryons and its effects on the galaxy evolution process.
We also introduce a novel set of modified DMO simulations which can reasonably replicate the effects of fully realized hydrodynamics on the DM halo while simplifying the analysis and lowering the computational cost.
We find that baryons change the thermal structure of the central region of the halo to a greater extent than the SIDM scatterings for MW-like galaxies.
Additionally, we find that the new thermal structure of the MW-like halo causes SIDM to create cuspier central densities rather than cores because the SIDM scatterings remove the thermal support by transferring heat away from the center of the galaxy.
We find that this effect, caused by baryon contraction, begins to affect galaxies with a stellar mass of $10^8$ M$_\odot$ and increases in strength to the MW-mass scale.
This implies that any simulations used to constrain the SIDM cross sections for galaxies with stellar masses between $10^8$ and at least $10^{11}$ M$_\odot$ will require baryons to make accurate predictions.
\end{abstract}

\begin{keywords}
methods: numerical - galaxies:halos - dark matter 
\end{keywords}


\section{Introduction}

The favored $\Lambda$CDM cosmology provides a simple and effective explanation for many features of our Universe.
These features range from large scale structure \citep{2014Hopkins} and halo formation \citep{2014Vogelsberger_nature, 2017Grand} to explaining the bullet cluster \citep{2006Clowe} and baryon acoustic oscillations \citep{2014Planck}.
However, discrepancies between simulations and observations motivate consideration for alternative dark matter (DM) models \citep{2017Bullock, 2020Vogelsberger}.
Some DM models have been ruled out from constraints placed from direct detection experiments in particle colliders and from astrophysical constraints.
However, a wide range of models still contain viable DM alternatives \citep{2018Boveia}.
The simplicity of the CDM model, especially when compared to the light sector, raises the question of whether these discrepancies are indicative of nuances missing from the CDM framework.

The most widely discussed problems with CDM lie in the population and characteristics of dwarf galaxies. These problems were first noted in N-body simulations, henceforth referred to as dark matter only (DMO) simulations. These simulations predicted many more satellites, large enough that they should contain stars, around Milky Way (MW) like galaxies than had been observed \citep[missing satellites;][]{1999Klypin}. Additionally, there were too many satellites with high central densities that do not have an observed counterpart \citep[too big to fail;][]{2011Boylan}. These satellites appear cored in observations, but remain cuspy in CDM simulations \citep[core-cusp;][]{1997Blok}. Finally, the satellite population in observations shows much more diversity in their rotation curves than those expected from simulations \citep[diversity;][]{2015Oman}.

There are two main classes of solutions that have been proposed to solve these problems in simulations, change the properties of the light sector or change the properties of the dark sector.
Modified Newtonian dynamics also poses another possible solution, however, we do not discuss this here and instead point the reader to the review by \cite{2015McGaughA}.
Changing the properties of the light sector can occur through hydrodynamical simulations which augment DMO simulations by adding baryons that abide by the laws of fluid dynamics and allow for additional galaxy formation physics.
Adding baryons, in the form of gas, stars, and black holes, allows for a more realistic galaxy formation process, adds additional observables to compare to observations, and can have a dynamical impact on the DM \citep{2015Chan, 2020Callingham}.
These additional features allow hydrodynamical simulations to improve upon the initial predictions from DMO simulations and have helped alleviate some of the small-scale problems.
These simulations have shown that not all DM halos will be populated with a luminous galaxy \citep{2018Simpson}, baryons can induce core formation \citep{2015Chan}, and the number of satellites around MW-like galaxies can vary between halos \citep{2021Engler}.
These simulations show progress toward alleviating the missing satellite and core-cusp problems.
However, the addition of baryons in simulations without bursty feedback has not helped to alleviate the core-cusp problem, and has even increased the central densities in a wide range of galaxies \citep{2018Lovell}.
Other baryonic prescriptions, namely those which include burstier stellar feedback, have alleviated this problem in bright dwarf galaxies ($10^7$ - $10^9$ M$_\odot$), however the problem may still remain in smaller dwarfs \citep{2019Fitts}.
Together, the addition of baryons with explicit feedback models provide a promising outlook on the small-scale problems in $\Lambda$CDM.

While hydrodynamic simulations have begun to alleviate some of the small-scale tensions in $\Lambda$CDM, discussion continues on if baryons alone can solve these problems. 
In dwarf galaxies, feedback becomes inefficient at lower mass scales and does not form cores under $10^6$ M$_\odot$ \citep{2015Onorbe,2017Fitts,2020Lazar}.
Additionally, changes to the star formation history in models with burstier feedback can lead to variations in otherwise similar halos \citep{2016Read,2016Sawala,2017Fitts}.
These variations can make it difficult to understand the details on how a particular physics model will affect the DM halo.

The impact of baryons depends on the feedback model implemented \citep{2015Chan,2018Lovell,2020Callingham, 2020Lazar}. Non-bursty feedback leads to strong adiabatic contraction \citep{1986Blumenthal} and bursty feedback can expand the halo and lead to core formation \citep{2014Brooks}. 
In non-bursty feedback models, baryons are more likely to condense in the center of their halo through an adiabatic process. 
In this case, the DM halo is unaffected by the baryons other than changes induced by the deeper potential of the centrally-concentrated baryons \citep{2020Callingham}. 
If the gas cools rapidly or is explosively expelled from the inner regions of the halo, which is common in models with burstier feedback, the adiabatic assumption can be violated. 
The properties of dark matter halos then depend on the frequency and severity of the bursty feedback.
The IllustrisTNG simulations feature a subgrid  model for stellar feedback.
This model allows gas to cool, but does not explicitly model the detailed structure of the ISM \citep{2018Pillepich,2018Hopkins}. \cite{2018Lovell} found that the strength of the DM contraction in the IllustrisTNG simulations varied with the mass of the galaxy and resulted in DM concentration in all galaxies between $10^8$ and $10^{12}$ M$_\odot$ with the largest contraction in MW-like galaxies.
The Auriga simulations employ a galaxy formation model similar to that of IllustrisTNG on MW-like galaxies and found the adiabatic assumption underestimated the contraction in their simulations by $\sim$8\% \citep{2020Callingham}. 
The FIRE simulations differ from IllustrisTNG-style physics by including explicit feedback routines \citep{2014Hopkins, 2015Muratov}. 
The explicit feedback routines aim to resolve star formation sites and deposit the feedback locally.
This feedback often leads to burstier star formation, especially at early times and in gas-rich environments.
\cite{2020Lazar} found that baryon contraction in MW-like galaxies often leads to an increase in DM density near the stellar half-light radius in the FIRE simulations. However, smaller galaxies ($10^8$ - $10^9$ M$_\odot$) showed less dense DM in this region due to stronger stellar feedback and the formation of larger cores \citep{2020Lazar}.

Alternative to modifying the baryon physics model used in the simulations, various modifications to DM have been proposed to alleviate some of the small-scale problems present in $\Lambda$CDM. The most common modifications to DM include warm dark matter \citep[WDM;][]{2009Kusenko} and self interacting dark matter \citep[SIDM;][]{1992Machacek, 2000Spergel}. 
WDM dampens the small-scale structure found in CDM simulations that led to the missing satellite problem. 
The model allows for greater DM thermal velocities at early times which allows the DM to overcome small perturbations in the initial density field. 
The particle streaming velocity and thus the scales over which power is suppressed are dependent on the model chosen. 
SIDM allows for interactions between dark matter particles which modify the internal structure of DM halos. These interactions increase the velocity dispersion near the center of the halo and can increase the efficiency of heat transfer within the halo \citep{2012Vogelsberger,2013Rocha,2016Vogelsberger}. 
Both of these DM models show promise in alleviating the small-scale problems in $\Lambda$CDM; however, most (but not all, as discussed below) of the results come from DMO simulation with few simulations including both baryons and modified DM.

DMO simulations remain a powerful tool astronomers can use to simulate a wide range of systems at a low computational cost. 
They are especially useful to test different properties of dark matter, such as warm dark matter \citep{2011Polisensky, 2014Lovell}, self-interacting dark matter \citep{2016Vogelsberger,2018Tulin}, and fuzzy dark matter \citep{2014Schive,2017Hui}, since the parameter space for these models is large. 
However, including baryons in these simulations provides a more accurate result and can significantly change the galaxy properties and formation process \citep{2015Fry, 2017Kamada}.
Thus, it is important to understand the regimes in which DMO results will remain robust once baryons are included.

Self interacting dark matter (SIDM) provides one example where many DMO simulations have been performed, but not many baryonic simulations. SIDM has been shown to be a promising candidate to resolve some of the small scale problems apparent in cosmological simulations \citep{2016Vogelsberger, 2013Rocha}. Namely, the core-cusp and the TBTF problems. 
SIDM alleviates these problems by causing an effective heat transfer toward the center of the halo to form an iso-thermal core.

Some work investigating how baryons affect SIDM halos has been done at both the dwarf scales and at the MW-mass scale. 
At the dwarf scale, \cite{2014Vogelsberger} found little difference in the SIDM density profiles for $10^8$ - $10^9$ M$_\odot$ halos once non-bursty baryons were included. 
However, they did find substantial differences to the baryonic distributions in the inner 1kpc.
\cite{2017Robles} simulated smaller dwarf galaxies ($10^6$ - $10^7$ M$_\odot$) with baryon physics from the FIRE-2 model \citep{2014Hopkins}. They also found minimal differences once baryons are added to these smaller dwarf halos.
\cite{2021Shen} and \cite{2022Shen} simulated a suite of dwarf galaxies ($10^5$ - $10^9$ M$_\odot$) with the FIRE-2 physics model and dissipative SIDM models.
Unlike the elastic SIDM models, \cite{2021Shen} found that DM energy dissipation can dominate over baryonic feedback for their fiducial models in dwarf galaxies.
Thus far, two simulation suites have been presented that include SIDM and baryons at the MW mass scale \citep{2021Sameie, 2021Shen}. 
There has been no analysis presented on the MW-like galaxies in \cite{2021Shen} so we do not discuss results on these galaxies here.
In the simulations presented in \cite{2021Sameie},  with the FIRE physics model, the SIDM halos had a greater response to the presence of baryons than the CDM halos which led to denser central regions. 
At larger mass scales, \cite{2019Robertson} found the cores that were produced by SIDM were diminished once baryons were added in $10^{14}$ - $10^{15}$ M$_\odot$ halos. They also found that the addition of baryons relaxed the constraints on SIDM cross sections at this scale.

In this paper we extend previous work by addressing the mechanism in which baryons affect SIDM halos.
Our goal is to understand where SIDM predictions are most affected by introducing baryons.
To this end, we introduce an idealized model which changes the potential without baryons and study how these results change from DMO and hydrodynamic simulations.
This paper is organized as follows. In Section \ref{sec:methods}, we outline the methods used in this paper which include the introduction of our simulation suite and a description of the SIDM model we simulate.
In Section \ref{sec:results}, we present the results from the simulations, focusing on how the addition of baryons and SIDM change t.he predictions about MW-like galaxies from other simulations.
In Section \ref{sec:discussion}, we discuss our results.
Finally, in Section \ref{sec:conclustion} we conclude.

\section{Methods}
\label{sec:methods}

In this paper we introduce six simulations of the same MW-like galaxy with different physics regimes.
Each simulation is a zoom-in simulation where the MW halo and all of its satellites are contained in a high-resolution region.
The volume outside this region is simulated at a lower resolution to speed up computation without greatly affecting the target of this investigation.
The halo used in this paper matches the halo used in \cite{2016Vogelsberger} which was chosen from a periodic 100 cMpc dark matter only (DMO) simulation with a DM mass resolution of $7.8 \times 10^7$ h$^{-1}$ M$_\odot$ and a spatial resolution of 2 h$^{-1}$ kpc. 
The halo was selected randomly from the halos between masses $1.58 \times 10^{12}$ and $1.61 \times 10^{12}$ M$_\odot$ and those which do not have another halo greater than 50\% their mass within 2 h$^{-1}$ Mpc. 
We note that this mass and isolation criteria mean this is a MW-sized galaxy but not a local group analog.

The DM mass and baryon mass resolution in our high-res simulations are $1.9 \times 10^5$ M$_\odot$ and $2.3 \times 10^4$ M$_\odot$, respectively.
The spatial resolution from gravitational softening is 158 pc at redshift 0 with a minimum cell size of 18 pc. 
The zoom-in initial conditions were formed at redshift 127 with \textsc{music} \citep{2011Hahn}. 
We adopt cosmological parameters from \cite{2016Planck}, where $\Omega_m$ = 0.301712, $\Omega_b$ = 0.046026, $\Omega_\Lambda = 0.698288$, and H$_0$ = 100 h km s$^{-1}$ such that h = 0.6909.

We use the \textsc{arepo} code \citep{2010Springel} for these simulations and employ the physics from the IllustrisTNG model \citep{2018Pillepicha, 2018Weinberger}.
The IllustrisTNG model improves upon the Illustis model \citep{2014VogelsbergerA, 2014Torrey} and has been tested thoroughly to show that it can generally reproduce realistic galaxies \citep{2018Pillepichb,2018Springel}.
The physics model includes star formation and feedback with galactic outflows, black hole formation and feedback, and gas enrichment and cooling \citep{2018Pillepich}. Section \ref{sec:realGal} compares the general properties of the galaxies in this paper with observations.

The six simulations can be categorized in three groups, each with a CDM simulation and a SIDM simulation. 
The three groups, which will be discussed further in this section consist of: DMO simulations, modified DMO simulations (described below), and hydrodynamical (baryonic) simulations. 
The CDM simulations are used as the baseline for comparison in each group to understand the effects of SIDM.
The DMO simulations are used as the baseline to understand how baryons change the predictions from other simulations. And the modified DMO simultions are used to disentangle the various baryonic effects on the DM halo.

The model for the self interacting dark matter (SIDM) is identical to the one described in \cite{2012Vogelsberger} and implemented in \cite{2016Vogelsberger}. 
Generally, this approach uses a Monte Carlo algorithm to give each particle a probability of scattering with its $32 \pm 5$ nearest neighbors. 
This number was chosen empirically to run the neighbor search efficiently without greatly affecting the results \citep{2012Vogelsberger}.
Simulation time steps are chosen such that at most one scattering occurs in a single time step or determined from dynamical constraints. In the elastic case, which we simulate here, once two particles scatter, their speed is taken to be equal in the center of mass frame and their directions are chosen at random. This technique conserves energy and linear momentum, but not angular momentum.

\begin{figure}
	\includegraphics[width=\columnwidth]{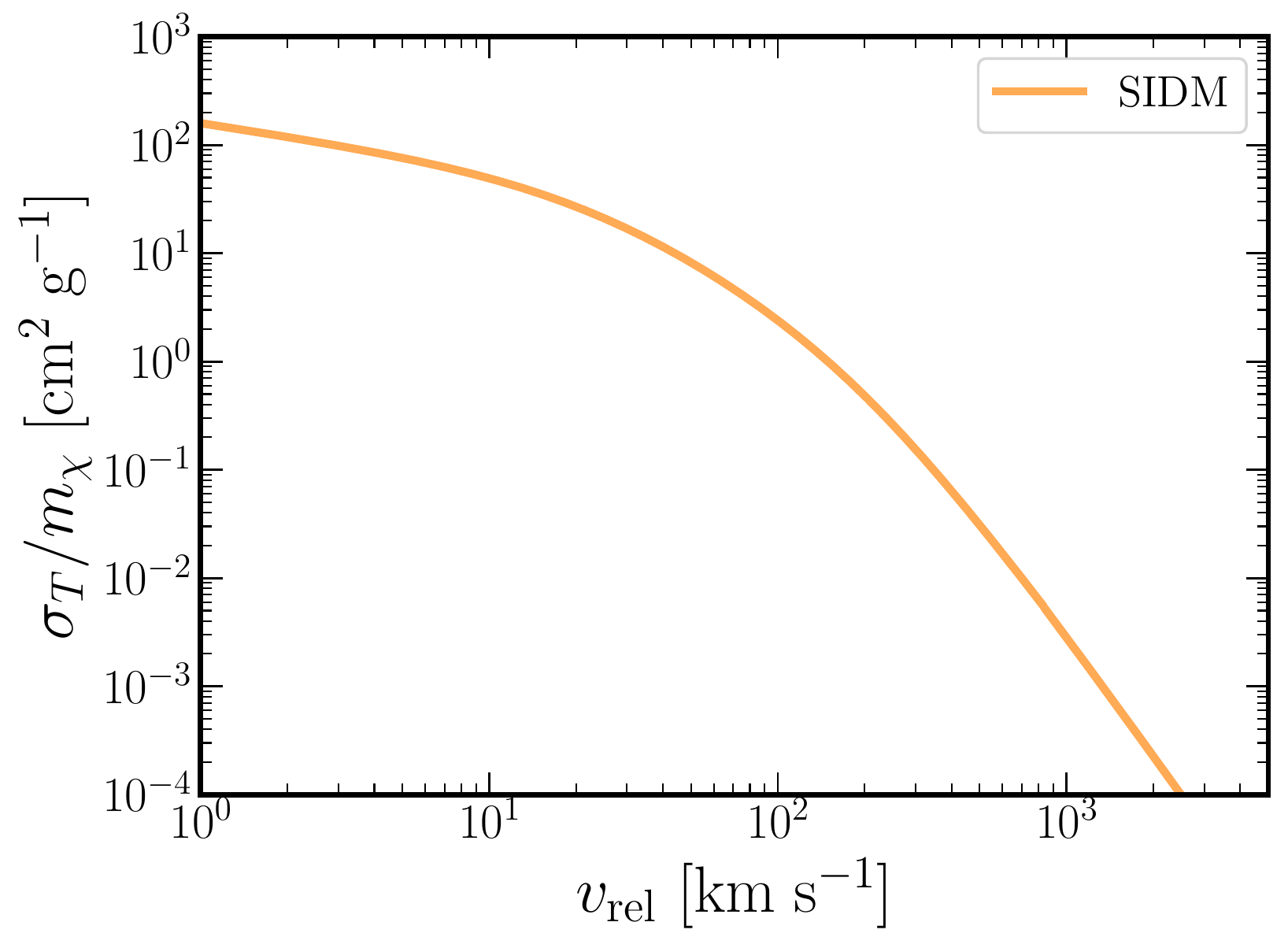}
    \caption{The one velocity dependent SIDM cross section we use in this paper. The plot is reproduced in a reduced form from \citet{2016Vogelsberger}. }
    \label{fig:cross}
\end{figure}

In this paper, we present one velocity-dependent cross section with elastic collisions taken from the ETHOS framework \citep{2016Cyr-Racine}. 
The ETHOS framework creates a mapping from the dark matter particle properties to the linear matter power spectrum and the DM transfer cross sections relevant in cosmological simulations.
We do not present simulations of the entire ETHOS framework, including both the velocity dependent cross sections and the modified initial matter power spectrum.
Instead, we present simulations with only the velocity-dependent self interactions to decouple the effects from the two and to focus on the changes to the internal halo structure from SIDM. 

The SIDM model we use corresponds to ETHOS-3 in \cite{2016Vogelsberger} and will be referred to here as `SIDM'. For reference, the velocity-dependent cross section is shown in Figure \ref{fig:cross}. The cross section from ETHOS-3 was chosen for this paper because it has the largest average cross section and causes the largest changes to the internal structure of the halo of the four ETHOS models presented in \citep{2016Vogelsberger}. 

\begin{figure}
	\includegraphics[width=\columnwidth]{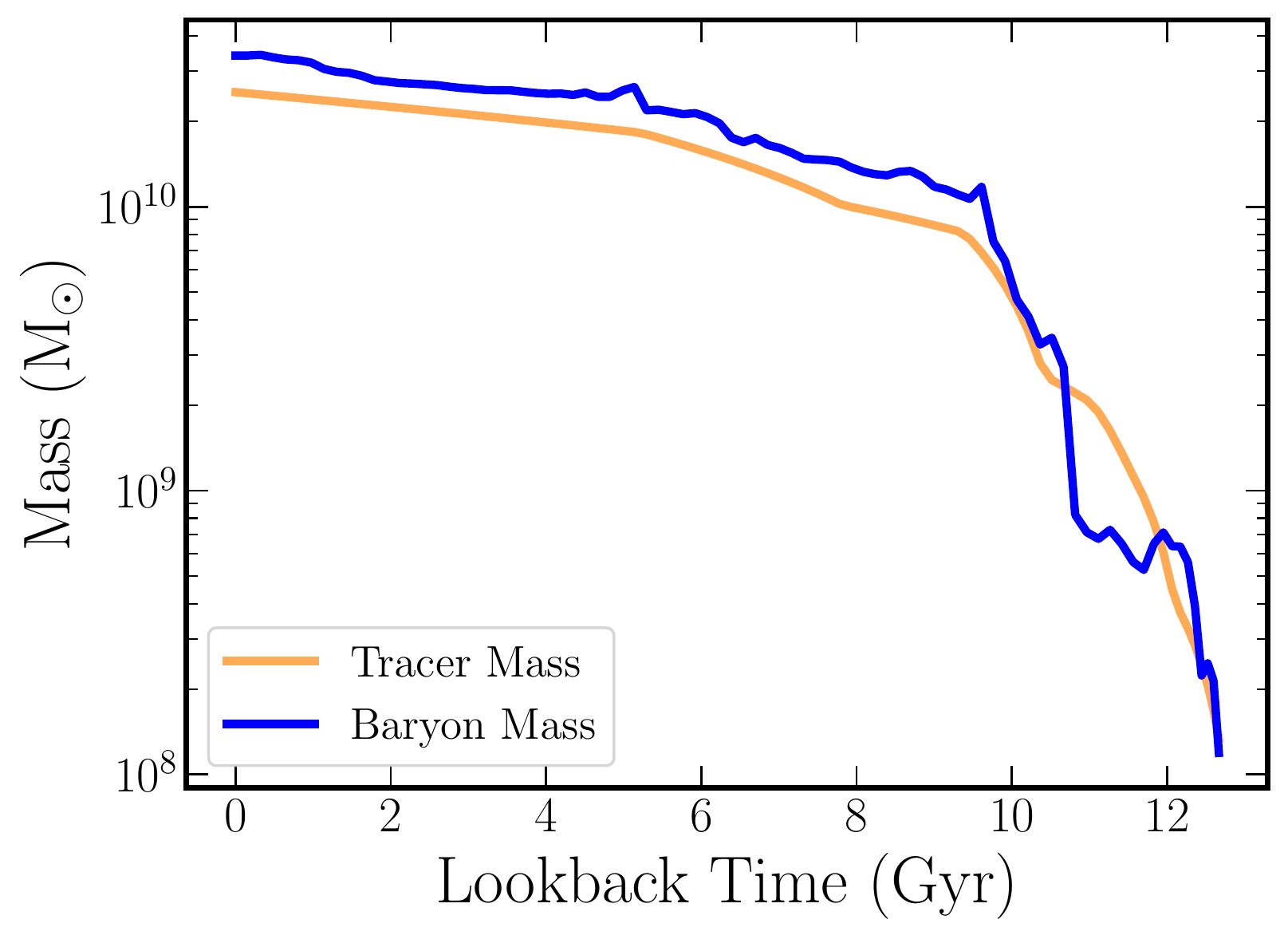}
    \caption{This figure shows the mass of the massive tracer used in the modified DMO simulations and the mass in the central 5kpc of the baryonic CDM simulation over time.
    The massive particle grows smoother than the baryon mass and does not align exactly.
    Overall, the two masses grow at very similar rates over 12 Gyr.}
    \label{fig:BHvsB}
\end{figure}

In this paper, we aim to understand how baryons affect DM halos in SIDM simulations.
This will prove to be a difficult task as baryons undergo complex dynamics and their effects cannot always be intuitively understood.
To aid in our understating, we have developed a new set of simulations that simplify the baryonic evolution and allow us to better understand how baryons couple to the SIDM halo. 
These simulations are similar to the DMO simulations in that they do not include any baryonic matter, but they are modified to include a massive inactive particle at the center of the central halo. 
Henceforth, these simulations will be referred to as `modified DMO' simulations.
The massive particle is seeded into the central halo at redshift 6 with a seed mass of $10^8$ $M_\odot$. The mass of the particle is updated artificially at each time step to match the baryonic mass in the inner 5 kpc of our fiducial baryonic central halo. A plot of the particle mass vs time in modified DMO simulations and a plot of the inner 5 kpc baryonic mass in our fiducial baryon simulation vs time are shown in Figure \ref{fig:BHvsB}.
While the mass of the tracer matches the baryonic mass reasonably, the tracer does not match the shape of the baryonic potential.
Instead of being concentrated to a disc which spans $\sim$10 kpc, the massive tracer is a constant density sphere with a radius of 1 kpc surrounded by a sphere of decreasing density out to 2.8 kpc.
As will be discussed later in Section \ref{sec:modified}, we find that this setup reasonably recreates the effects of baryons on the DM halo.
These simulations include the same resolution levels as the DMO and baryonic simulations discussed previously, one set includes SIDM and the other does not. The initial conditions are the same as the ones that are used in the DMO and baryonic simulations.

An important note is that mass conservation is not maintained in the modified DMO simulations. However, the deviation introduced by the single additional particle creates a $3 \times 10^{-5}$\% deviation from conservation and is not expected to have a large effect on the simulation outside the host halo discussed here.

\section{Results}
\label{sec:results}

\subsection{Impact of Model Variations}
\label{sec:realGal}

\begin{figure*}
	\includegraphics[width=.49\textwidth]{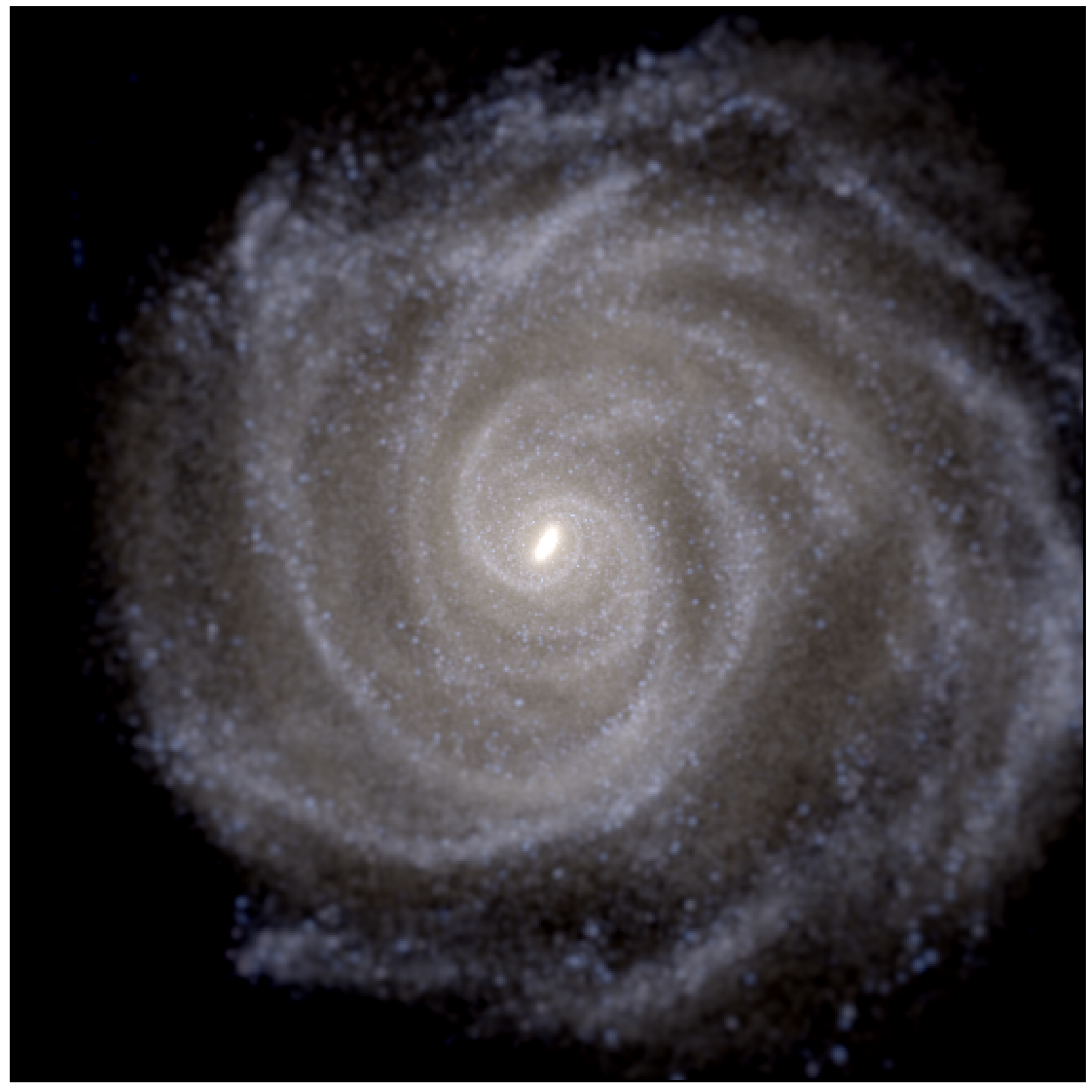}
	\includegraphics[width=.49\textwidth]{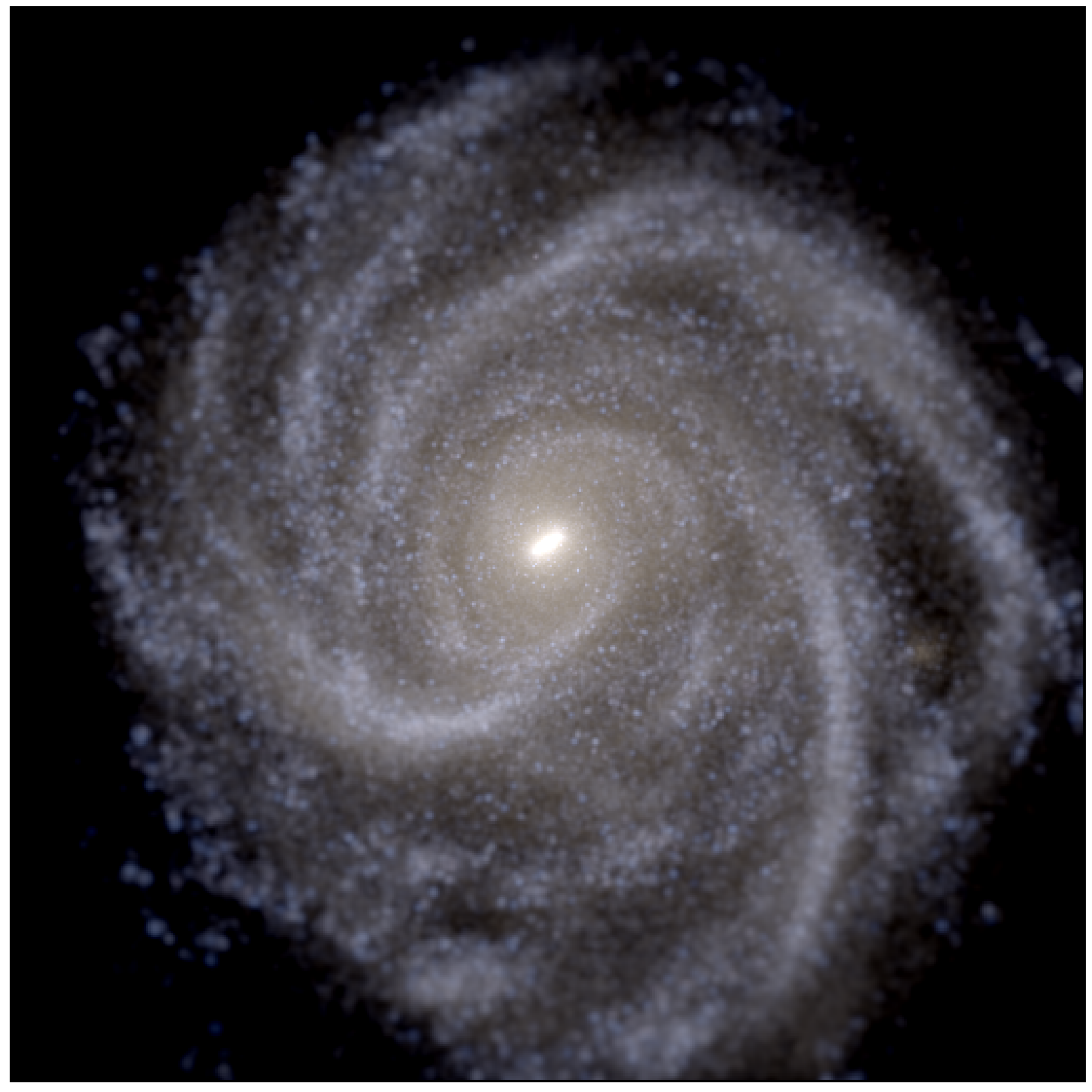}
	\includegraphics[width=.49\textwidth]{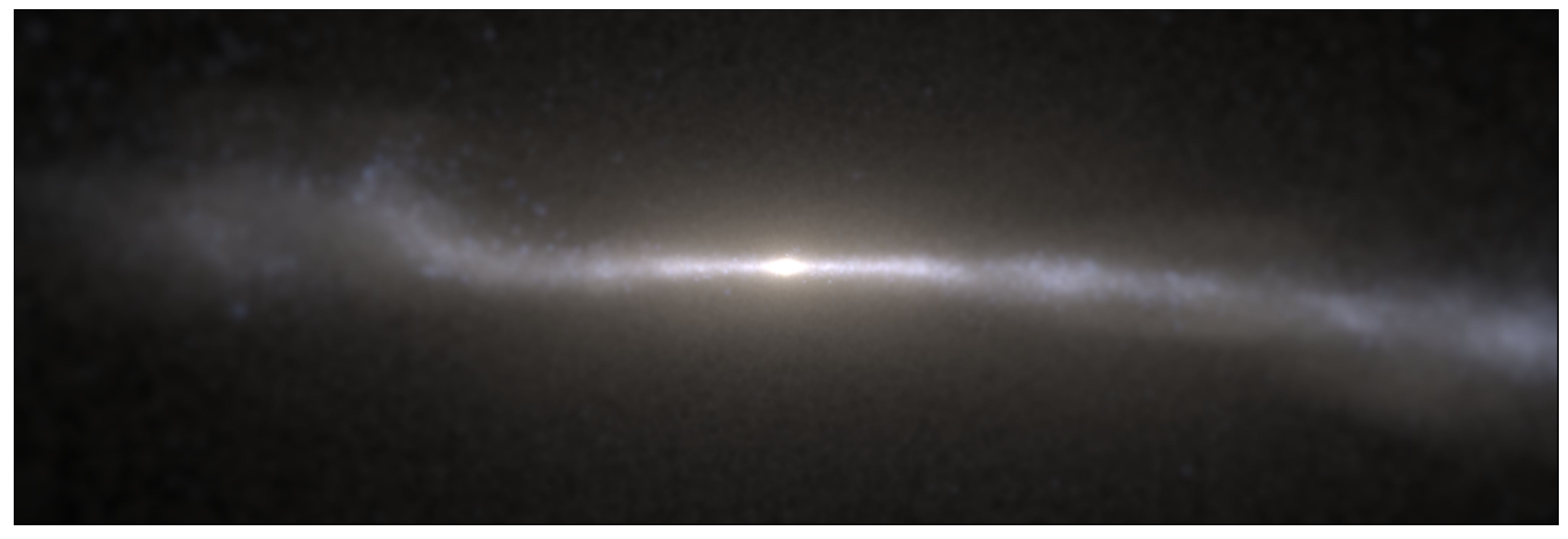}
	\includegraphics[width=.49\textwidth]{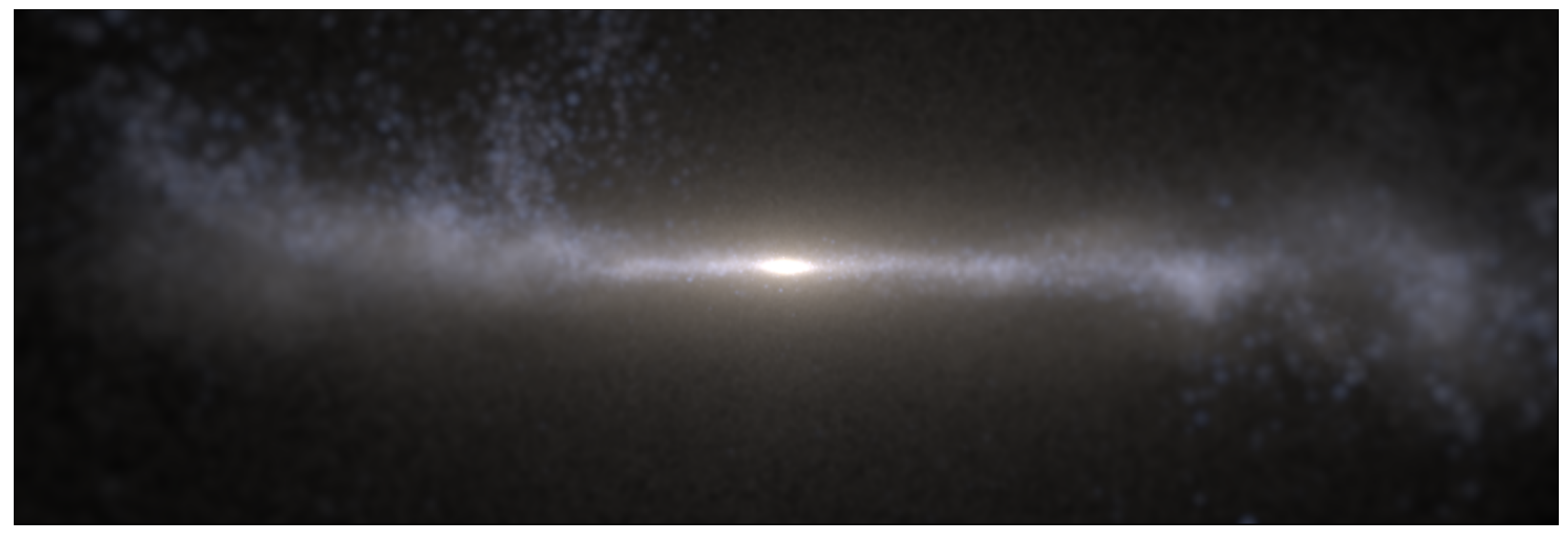}
    \caption{Composite g, r, i image of the central MW-like galaxy for the CDM (left) and SIDM (right) baryonic simulations. 
    Both simulations create galaxies with a clear spiral structure and thin disc. 
    The SIDM simulation shows signs of perturbations in the edge-on disc from a nearby satellite that will soon merge.}
    \label{fig:images}
\end{figure*}

First, we validate these models through a visual inspection and a comparison of the simulations to observations.
We present images of the central MW-like galaxy from our baryonic simulations at z=0 in Figure \ref{fig:images}. 
The face-on view of the galaxies was taken such that the angular momentum vector of the stars within 10 kpc of the center align with the vertical direction of the simulation box.
The edge-on view is the same as the face-on view but rotated 90 degrees.
The image is composed of the luminosities from the Sloan g, r, i filters for the blue, green, red frames respectively.

Each image covers a 60x60x20 kpc region which depicts a disc-dominated central galaxy with a bright central region and a spiral structure of young stars.
The size, shape, and color of the disc structure in both galaxies are similar.
Both galaxies support a thin disc of young blue stars surrounded by a thicker disc of older red stars.
In both simulations, the depicted central galaxy has just started a minor merger with an infalling satellite that is not shown here.
The satellite's proximity has caused some vertical warping in both galaxies which is more prominent in the SIDM galaxy as the satellite is slightly closer to the central.

\begin{figure*}
	\includegraphics[width=.33\textwidth]{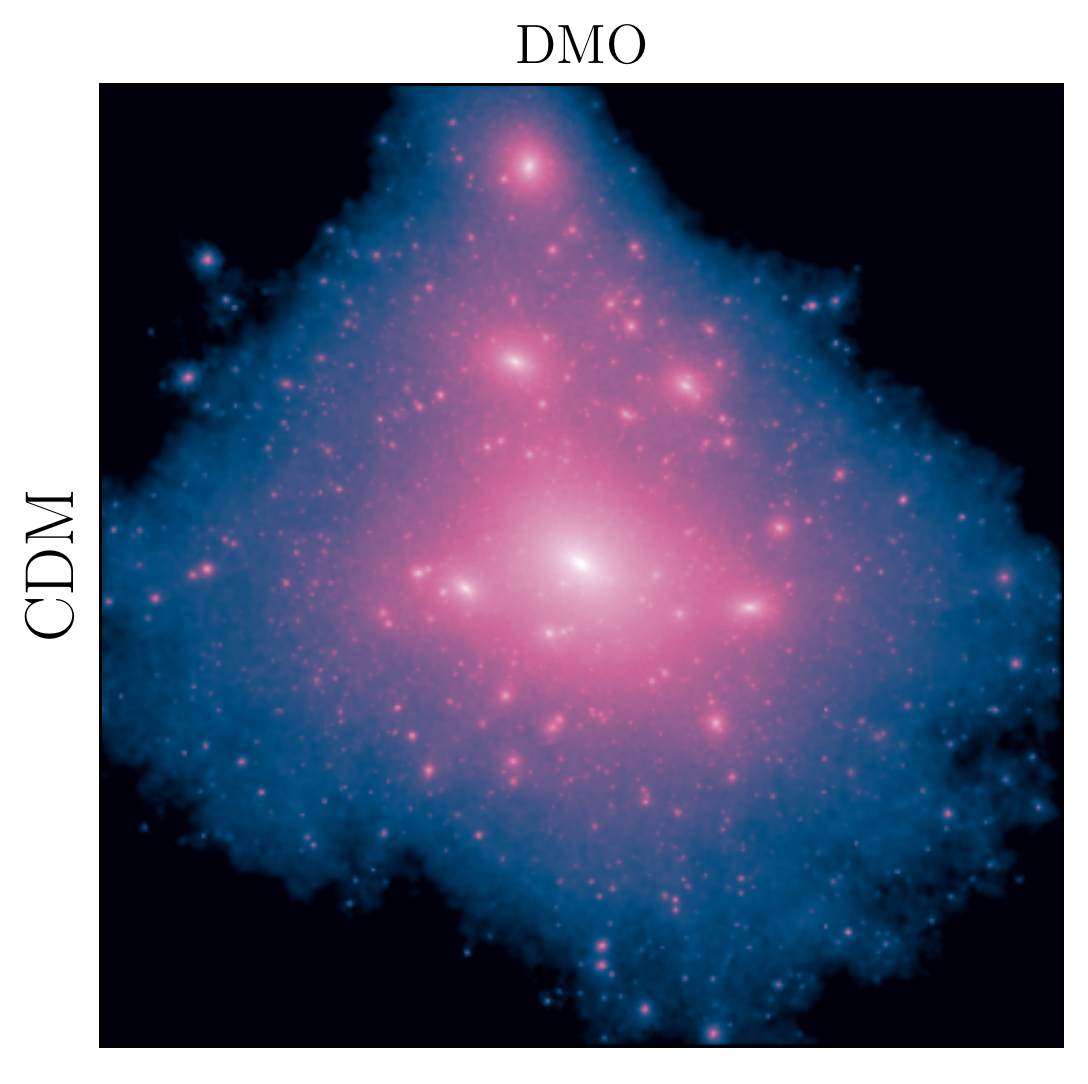}
	\includegraphics[width=.31\textwidth]{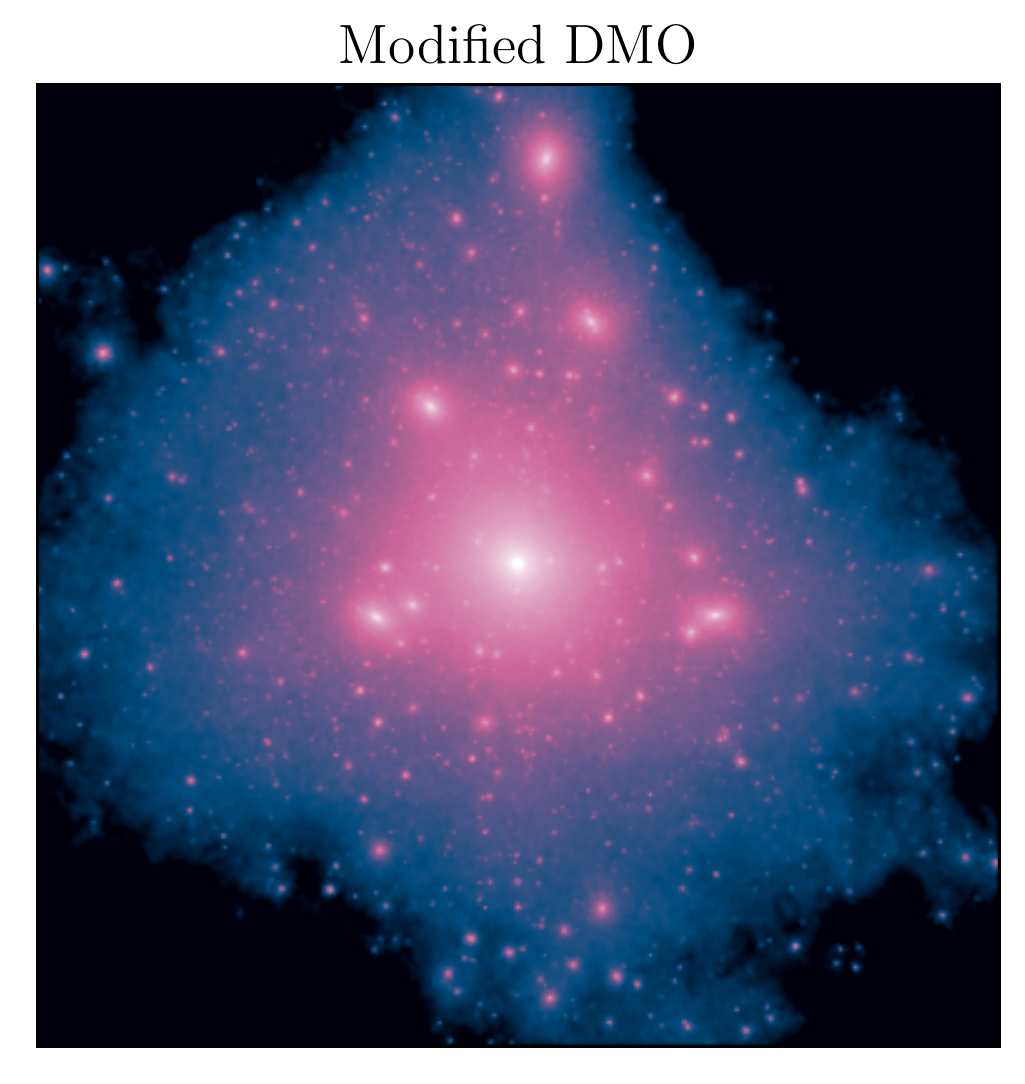}
	\includegraphics[width=.31\textwidth]{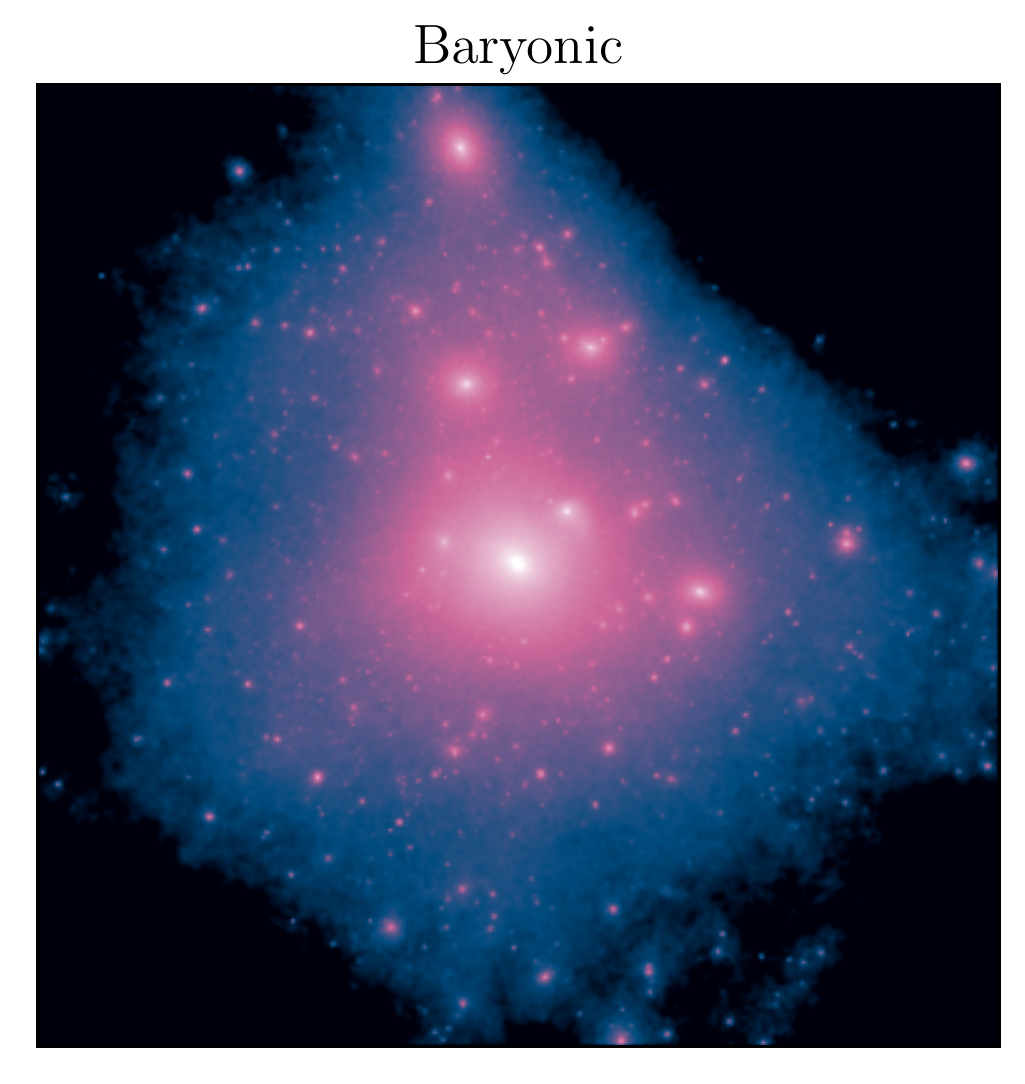}
	\includegraphics[width=.33\textwidth]{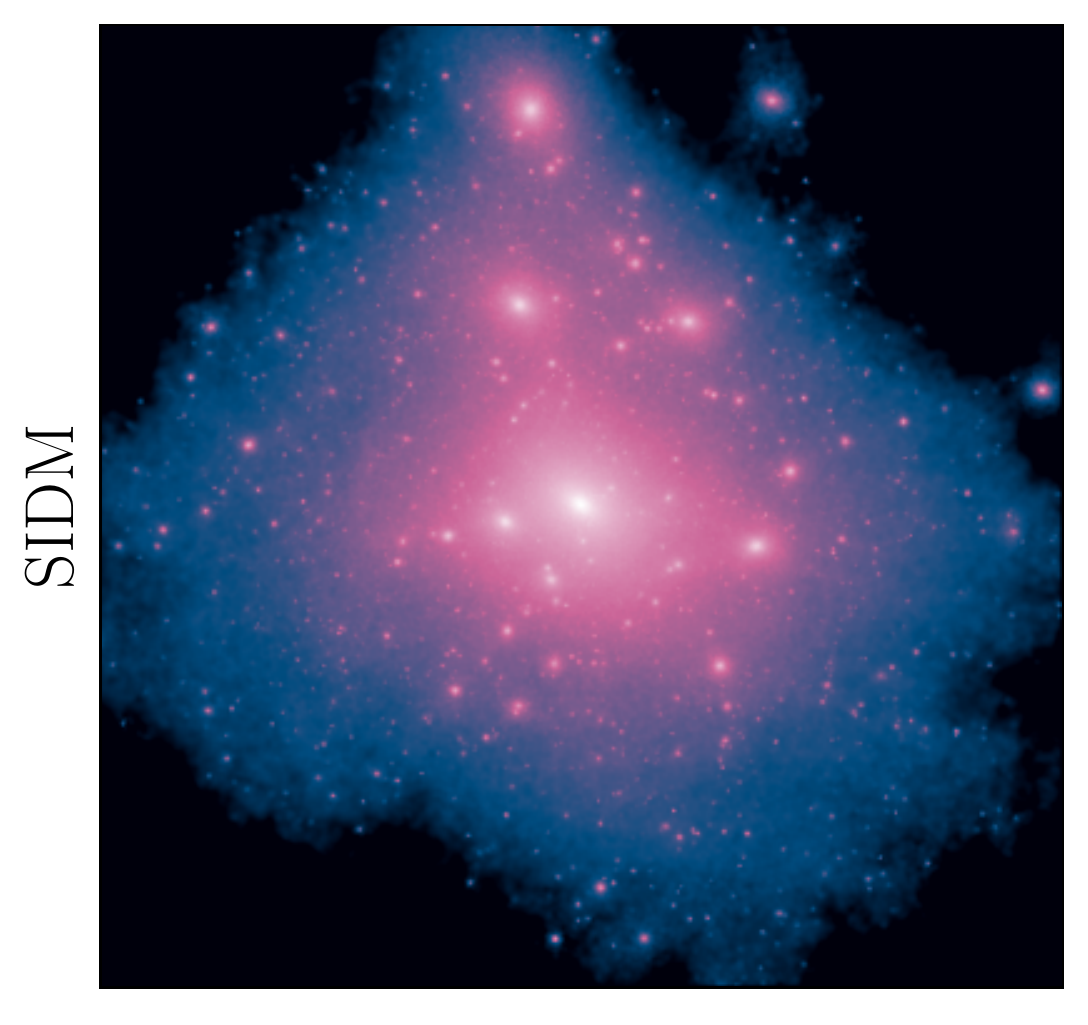}
	\includegraphics[width=.31\textwidth]{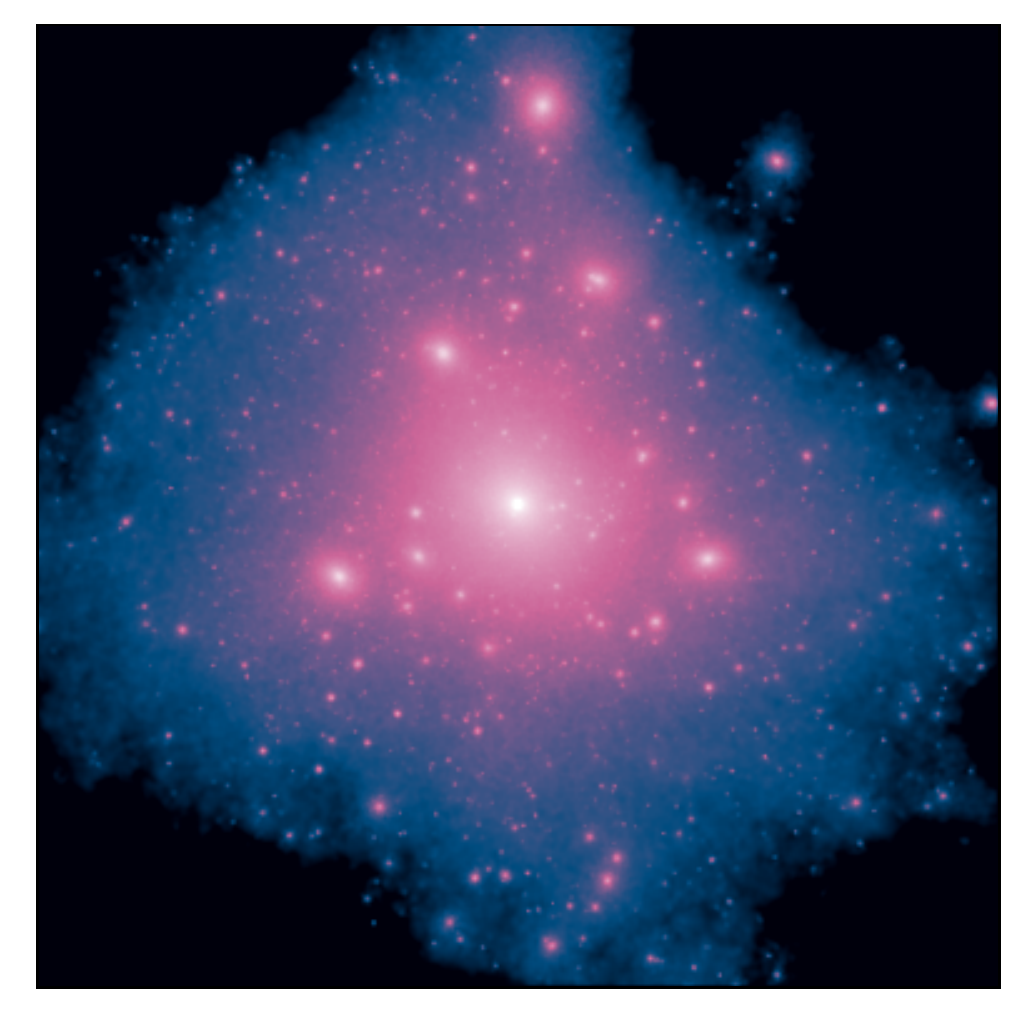}
	\includegraphics[width=.31\textwidth]{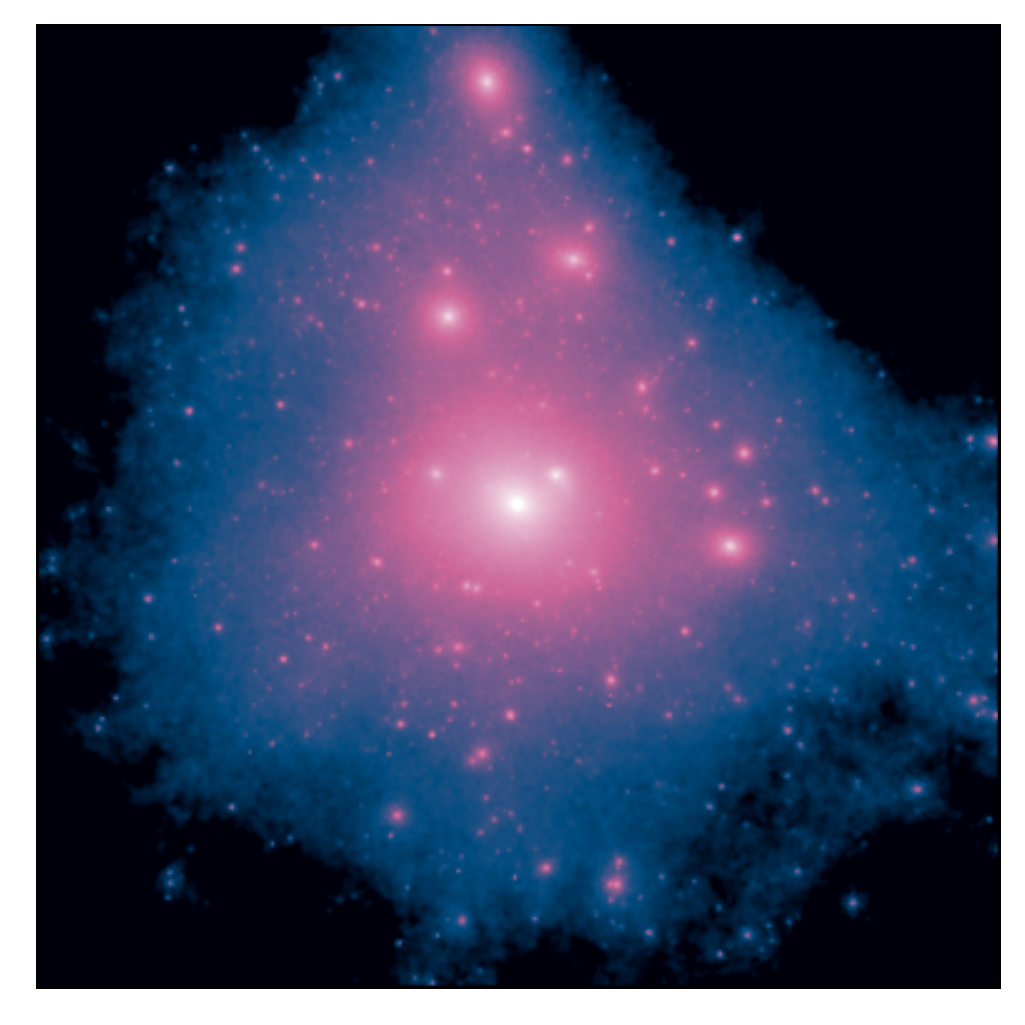}
    \caption{DM mass projections for the six fiducial simulations presented in this paper. 
    The top (bottom) row shows the CDM (SIDM) simulations.
    From left to right both rows show the DMO, modified DMO, and baryonic simulations.
    The projections are made from a 300x300x300 kpc box centered on the central MW-like halo.}
    \label{fig:images_DM}
\end{figure*}

Figure \ref{fig:images_DM} shows the DM halos for all six fiducial simulations.
The first (second) row shows the DM halos from the three CDM (SIDM) simulations.
From left to right, the simulations are DMO, modified DMO, and baryonic.
Qualitatively, there are no major differences between the six simulations other than some satellites being in different locations around the central halo.

\begin{figure*}
	\includegraphics[width=.33\textwidth]{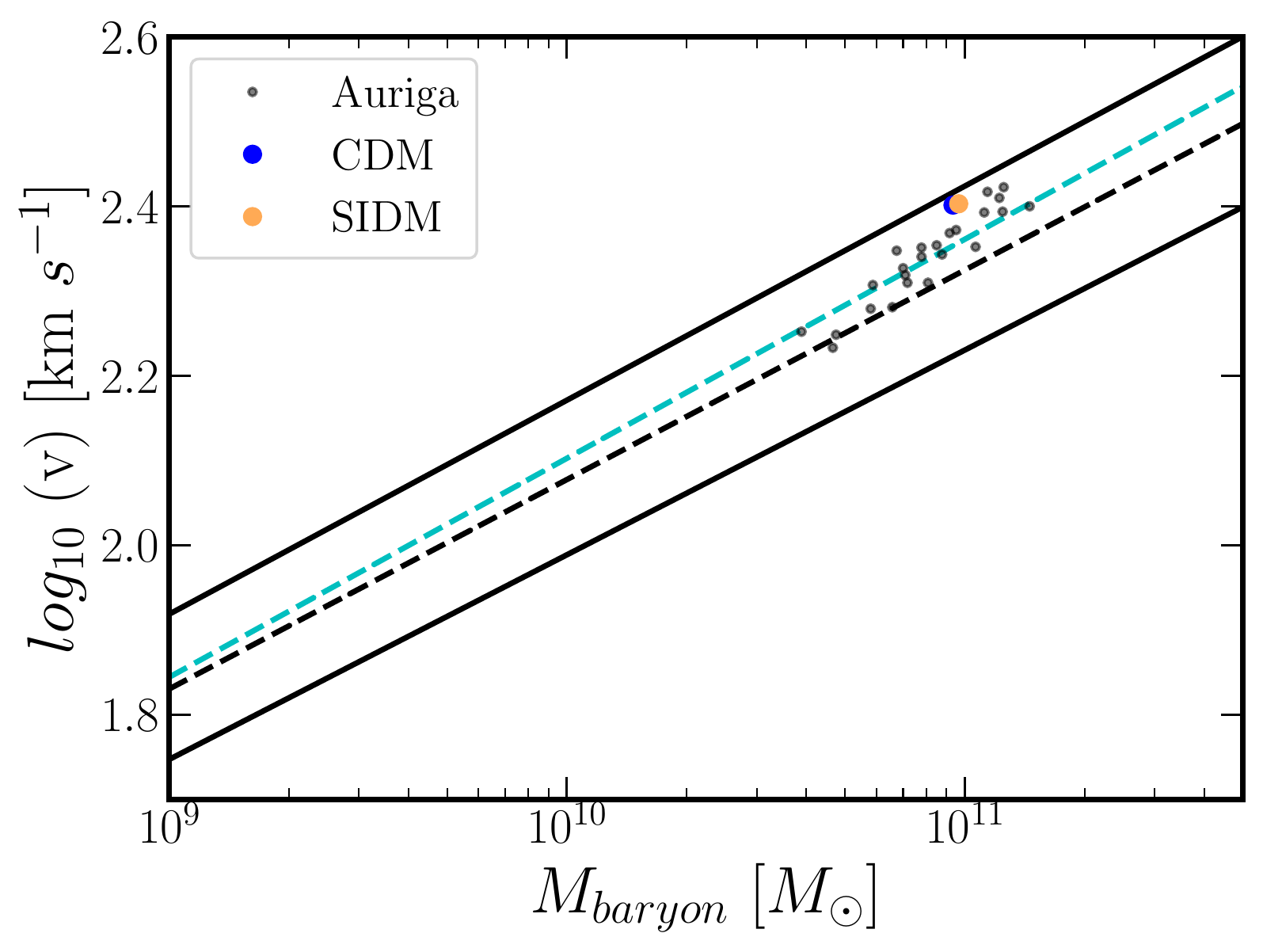}
	\includegraphics[width=.33\textwidth]{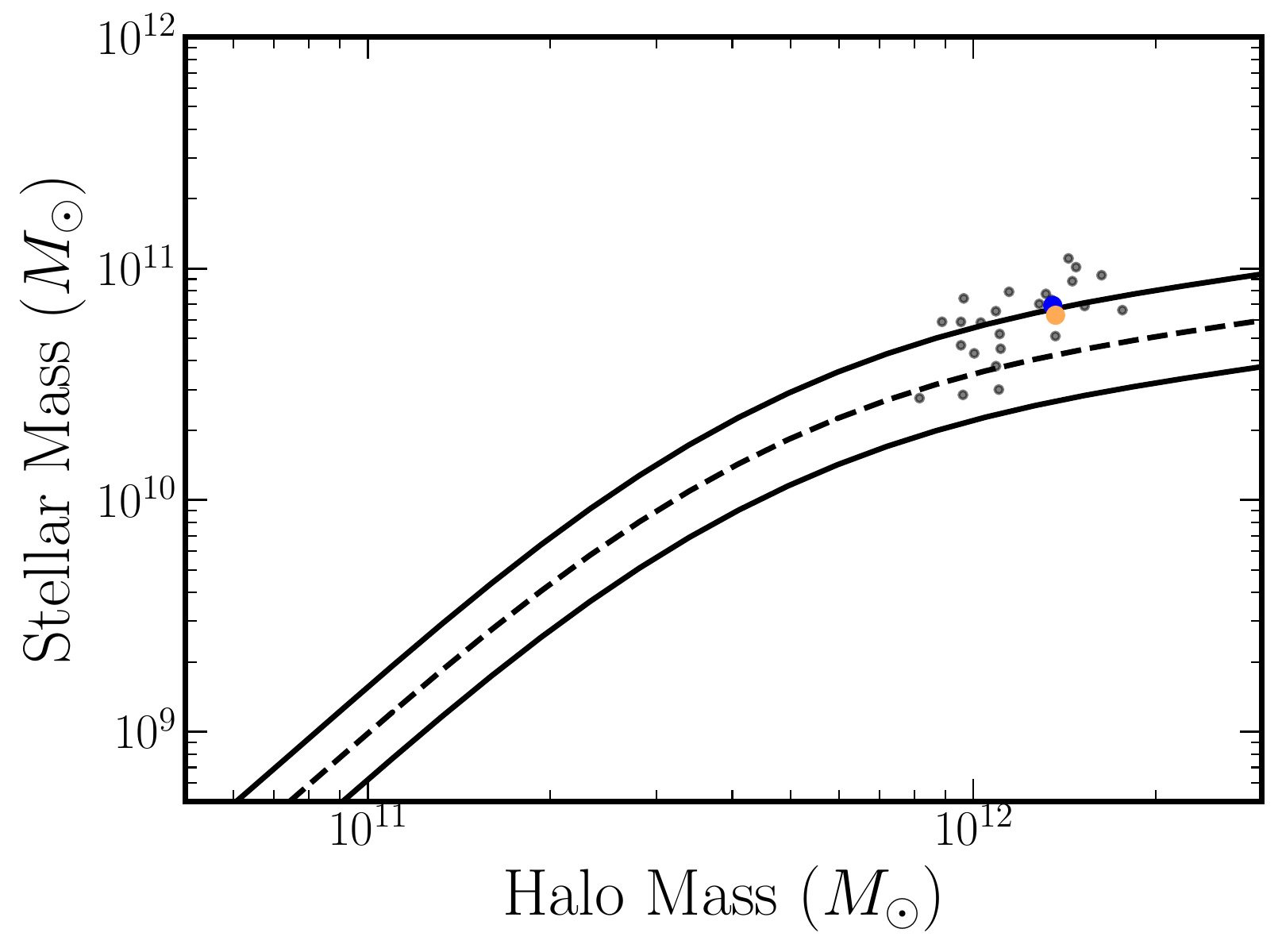}
	\includegraphics[width=.33\textwidth]{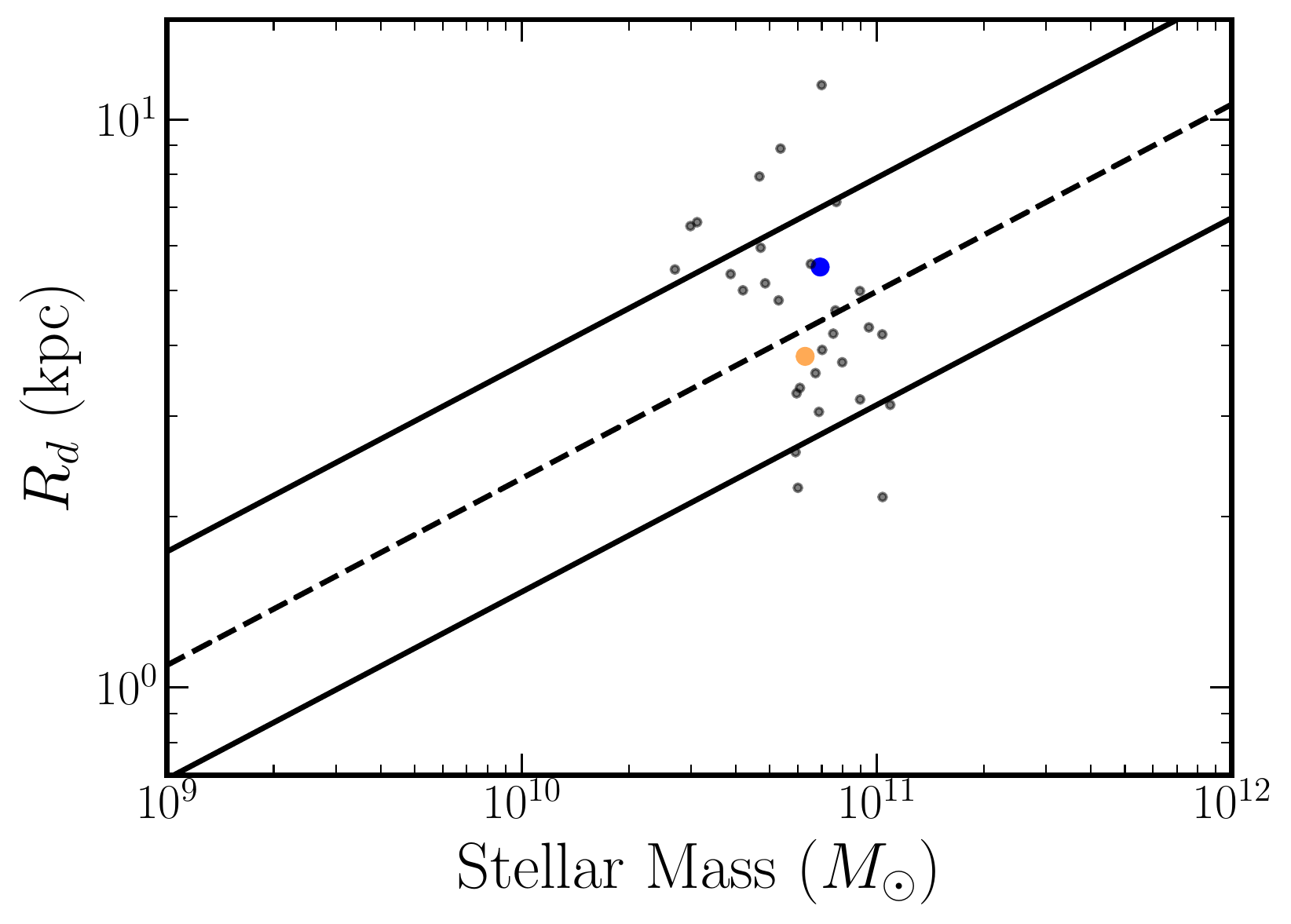}
    \caption{This figure shows a sample of observable properties from the two baryonic simulations to illustrate that these simulations form realistic galaxies.
    All plots use the same observational data as the Auriga simulations which also showed that their simulations produced reasonable galaxies \citep{2017Grand}.
    The results of the Auriga simulations are also presented here as gray points.
    (left) The Tully-Fisher Relation, observational data is taken from \citet{2015McGaugh}.
    (middle) The stellar mass - halo mass relation, observational data is taken from \citet{2013Moster}.
    (right) Disc radial scalelength vs stellar mass, observational data is taken from \citet{2009Gadotti}.
    For all three relations, which cover dynamics, size, and shape; both simulations fall within the observed scatter.}
    \label{fig:MW}
\end{figure*}

Next, we show that the galaxies produced in our baryonic simulations represent real MW-like galaxies through various observable relations in Figure \ref{fig:MW}. Shown from left to right, these relations include the baryonic Tully Fisher relation \citep{2000McGaugh}, the stellar-mass halo-mass relation \citep{2013Moster}, and the disc radial scalelength-stellar mass relation \citep{2009Gadotti}.

In each plot in Figure \ref{fig:MW}, the results from the simulations presented in this paper are shown as points and the lines show expected values from observed data. 
Our galaxies fall within the scatter of the observed relations for each comparison. 
Although, they are near the top of Tully-Fisher and stellar mass - halo mass relations. 
This result is consistent with the results from the Auriga simulations \citep{2017Grand} which also used these tests to determine if their simulations produced realistic galaxies. They also found that MW-like halos that were simulated with IllustrisTNG physics tend to fall near to upper end of the observed Tully-Fihser and stellar mass - halo mass relations.
The galaxies in our simulations fall near the center of the size-mass relation, but show the largest scatter of the three relations.
We adopt the same definition of scalelength as \cite{2017Grand} who analyzed the Auriga simulations: the inverse slope of an exponential fit to the outer 5kpc of the stellar surface density profile.
This result again agrees with the Auriga results which find a large spread in scalelength between different MW-like galaxies.
The two galaxies presented here are nominally the same galaxy, as they have the same initial conditions.
However, the large spread in the the Auriga radial scalelengths suggests that changes to the formation of the galaxy, here induced by including SIDM, strongly affect the radial scalelength.
The relations shown here provide a sample of observables we used to compare our simulation to observations.
We examined many other relations and found a similar level of agreement to the ones we show here over multiple redshifts including: stellar half-luminosity radius vs stellar mass, velocity dispersion profiles, SFR vs stellar mass and more. 

\begin{table}
    \label{tab:DM}
	\centering
	\caption{A comparison of values between the DMO and baryonic MW-like DM halo. The first column gives the name of the simulation. `dm' refers to dark matter only imsulations, `SIDM' refers to the self-interacting cross section shown in Figure \ref{fig:cross}. The second column gives the mass of the galaxy within 200 $\times$ the critical density in units of M$_\odot$. The third column gives the radius of a sphere centered on the halo with a mean density of 200 $\times$ the critical density in units of kpc. The fourth column gives the maximum circular velocity of the halo in units of km/s. The fifth column gives the radial distance from the center of the halo to V$_{\rm max}$ in units of kpc.}
	\label{tab:example_table}
	\begin{tabular}{l|cccc} 
		\hline
		Name & M$_{200}$ (M$_\odot$) & R$_{200}$ (kpc) & V$_{\rm max}$ (km/s) & R$_{\rm max}$ (kpc)\\
		\hline
		\hline 
		CDM      & 1.526 $\times 10^{12}$ & 239.6 & 219.3 & 12.22 \\
		SIDM     & 1.512 $\times 10^{12}$ & 238.9 & 223.0 & 0.584 \\
		CDM dm   & 1.586 $\times 10^{12}$ & 242.7 & 175.1 & 51.54 \\
		SIDM dm  & 1.593 $\times 10^{12}$ & 243.1 & 178.4 & 45.49 \\
		\hline
		\hline 
	\end{tabular}
\end{table}

In Table \ref{tab:DM} we list various parameters of the galaxies in the DMO simulations and compare them to the same values in the baryonic simulations. 
Given that these values in the DMO simulations are similar to those in the baryonic simulations, and that our baryonic simulations produce realistic galaxies, we conclude that our DMO simulations have produced generally realistic halos. There is some scatter between the DMO simulations and the baryonic simulations, but this is expected as the different physics regimens will produce slight differences in the halo properties.

\subsection{Impact of Baryons on Dark Matter}
\label{sec:bvsDMO}

\begin{figure*}
	\includegraphics[width=.335\textwidth]{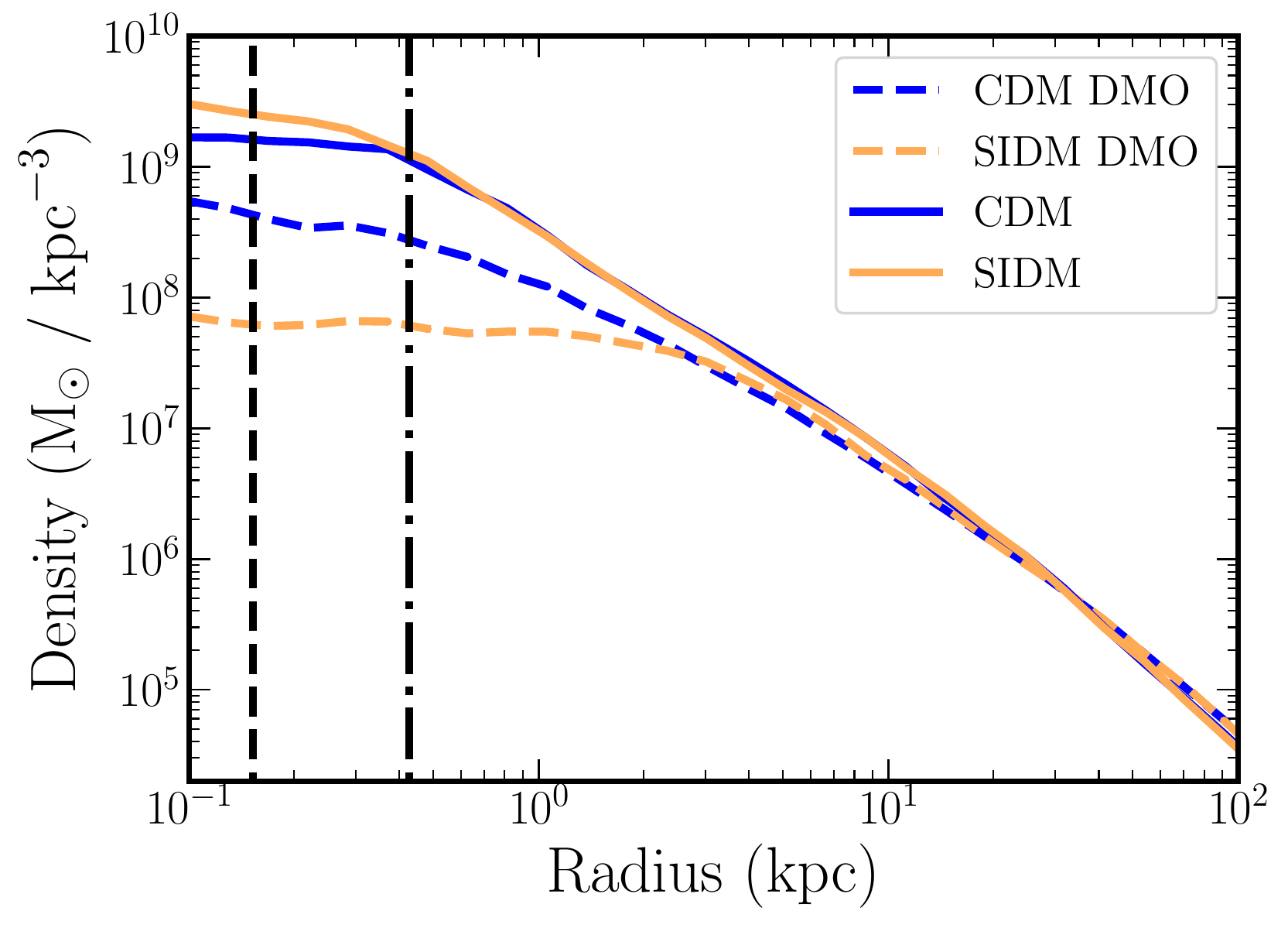}
	\includegraphics[width=.325\textwidth]{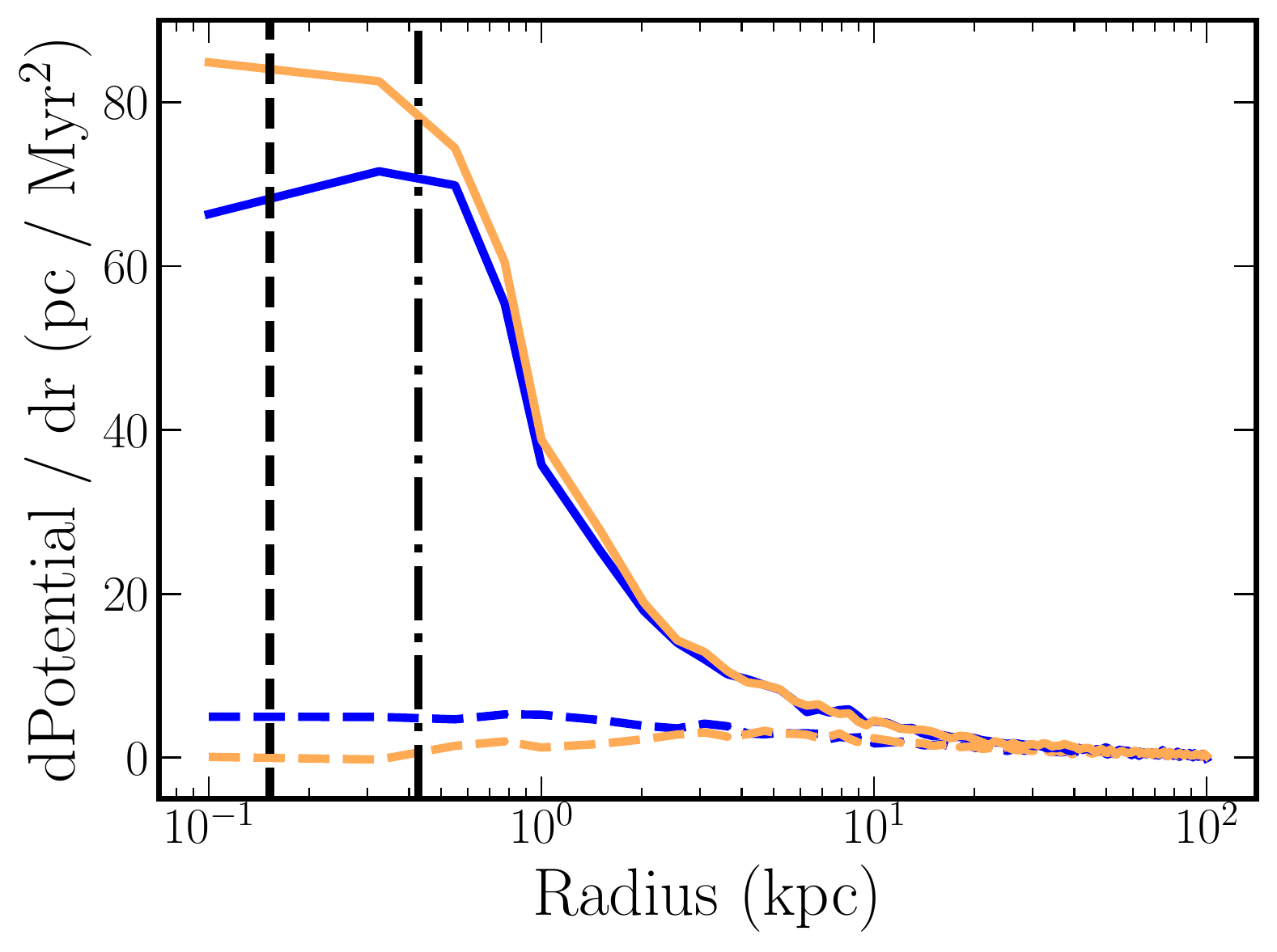}
	\includegraphics[width=.33\textwidth]{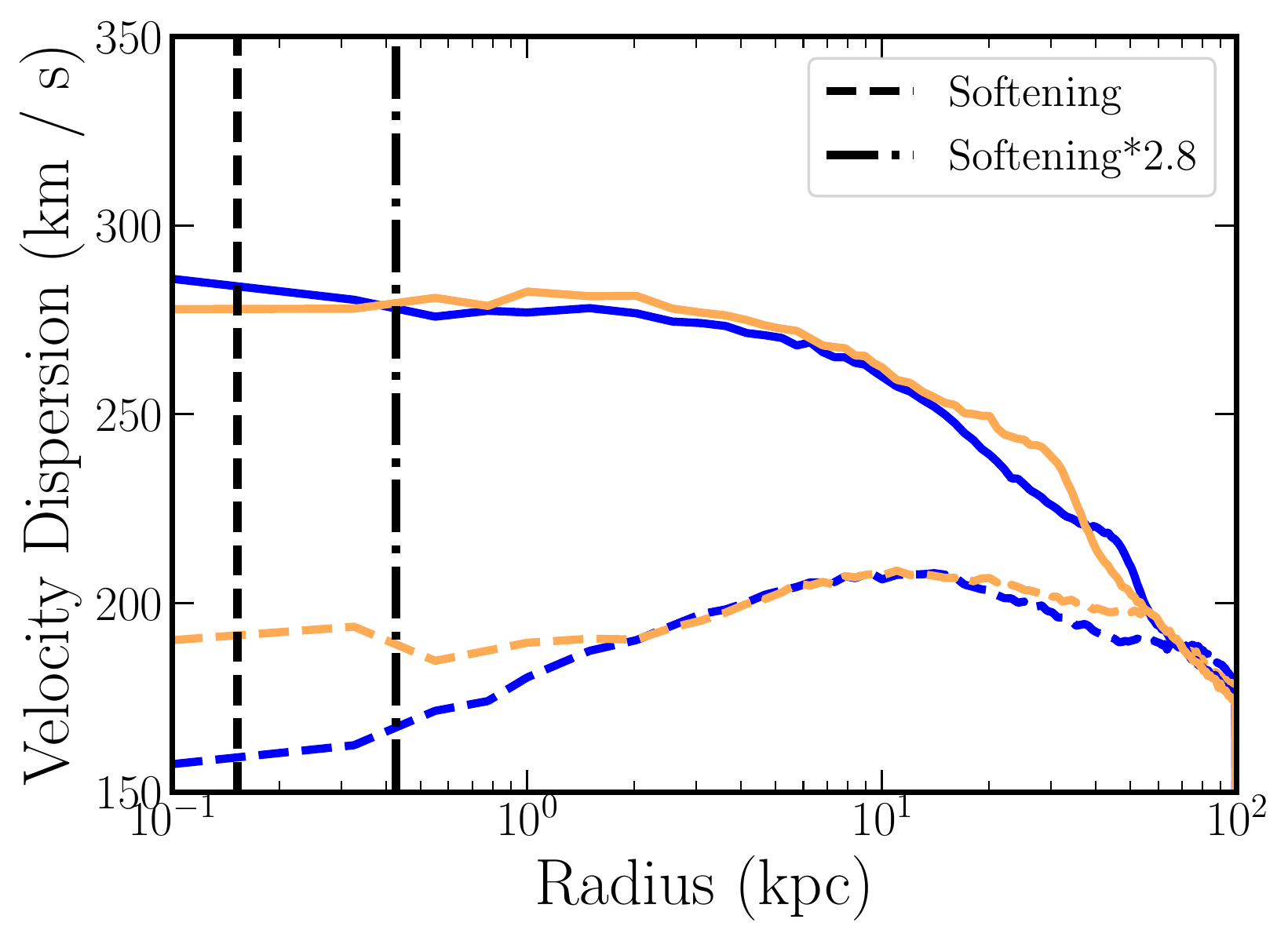}
    \caption{This figure shows how baryons affect the structure of the DM halo both with and without SIDM.
    `SIDM' refers to the simulation which uses the ETHOS-3 velocity dependant cross section, but not the modified initial matter power spectrum.
    The black lines shown in the middle panel depict the softening length for these simulations (dashed line) and 2.8 times the softening length (dot-dashed line).
    Inside 2.8 times the softening, numerical effects begin to affect the data.
    Inside the softening length, the data are greatly affected by the softening length.
    The left plot shows the DM density profiles for the central halo in each of the four simulations.
    The middle panel shows the slope of the DM potential profile.
    The right plot shows the velocity dispersion profiles.}
    \label{fig:DMvsB}
\end{figure*}

In this subsection, we investigate how adding baryons to these simulations affects the MW-like halo when SIDM is included by considering the spherically average DM density, potential slope, and velocity dispersion profiles.
The left plot in Figure \ref{fig:DMvsB} shows the spherically averaged radial DM density profile for both sets of CDM (blue) and SIDM (orange) simulations each with a DMO (dashed) and baryonic (solid) counterpart. The DM density profile is calculated by sampling the density field at 30 logarithmic-spaced locations with each location being the median of 900 density samples evenly spaced on a sphere. The SIDM DMO simulation shows the formation of a three kpc core while the CDM DMO simulation remains cuspy in this region.
The two profiles from the baryonic simulations agree very well up to 300 pc where the SIDM halo becomes slightly denser. 
There is approximately a half dex difference between the CDM DMO and baryonic profiles in the inner one kpc with convergence further out. While the density profiles of the DMO and baryonic simulations are nearly converged at 3 kpc, it is not until closer to 25 kpc where they actually converge. 

The spatial resolution from gravitational softening is shown as a dashed line at 153 pc. Inside this line, the shape of the potential, and therefore particle dynamics, are significantly influenced by gravitational softening.
This gravitational softening weakens particle interactions at short distances, which influences the recovered density, potential, and volocity dispersion profiles.
Additionally, a dot-dashed line is shown at 2.8 $\times$ softening length (428 pc) which marks the spline softening length where gravity returns to being fully Newtonian \citep{2020Ludlow}.

The middle panel in Figure \ref{fig:DMvsB} shows the first derivative of the potential with respect to radial distance of the baryonic simulations. 
The potential derivative is given rather than the potential to highlight how the addition of baryons increase the gravitational acceleration toward the center of the halo.
The first derivative of the potential is taken using a second-order numerical approximation. The potential derivatives agree between the DMO and baryonic simulations outside of 20 kpcs, but deviate greatly inside this region. The potential slope in the SIDM DMO simulation decreases slightly compared to the CDM DMO simulation due to the formation of a core. The CDM and SIDM baryonic simulation agree quite well up to a few hundred parsecs where there is a slight deviation. 

The right plot in Figure \ref{fig:DMvsB} shows the dark matter velocity dispersion profile for the baryonic and DMO CDM simulations. The velocity dispersion is calculated via
\begin{equation}
\label{eq:velDispersion}
    \sigma_{\rm v} = \sqrt{\mathrm{std}(v_x)^2 +
                           \mathrm{std}(v_y)^2 +
                           \mathrm{std}(v_z)^2}
\end{equation}
where $\mathrm{std}(v_i)$ is the standard deviation of the DM velocity distribution in the $x$, $y$, or $z$ direction.
Again, there is a difference between the DMO and baryonic velocity dispersion profiles within $\sim$60 kpc for both the CDM and SIDM simulations. 
The SIDM DMO simulation shows that the inner two kpc have formed an isothermal core, characterized by the flat velocity dispersion profile, consistent with the core formed in the DM density profile. Both the CDM and SIDM baryonic simulations show a larger thermalized region, extending to four or five kpc. However, this thermalized region does not correspond to a similar DM density core. 
Unlike the DM density and potential derivative profiles, the baryonic and DMO velocity dispersion profiles do not converge until 80-100 kpc, showing an even larger divergence once baryons are included.

\subsection{Impact of Baryons on Self-Interactions}
\label{sec:sidm}

\begin{figure*}
	\includegraphics[width=.33\textwidth]{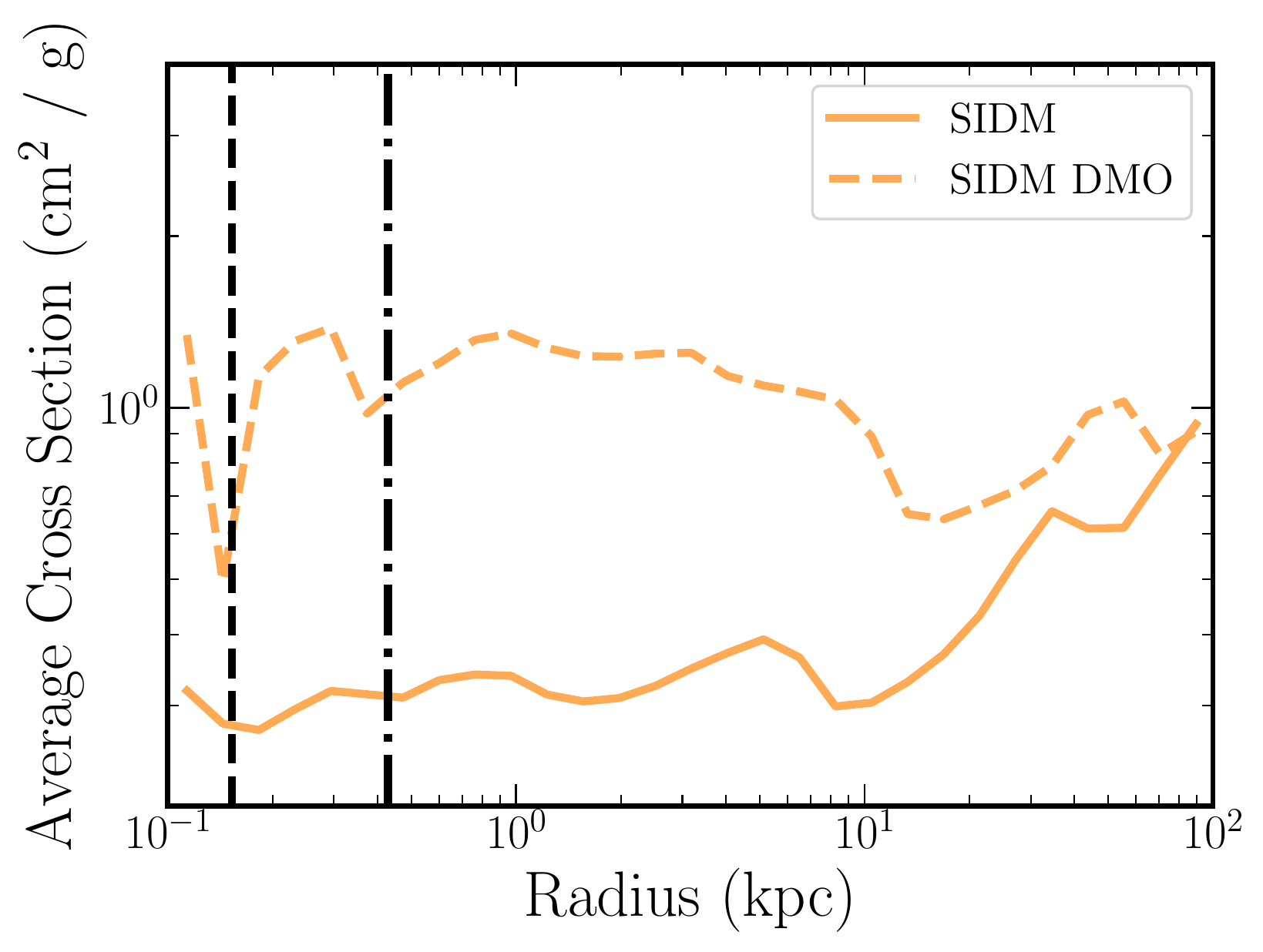}
	\includegraphics[width=.33\textwidth]{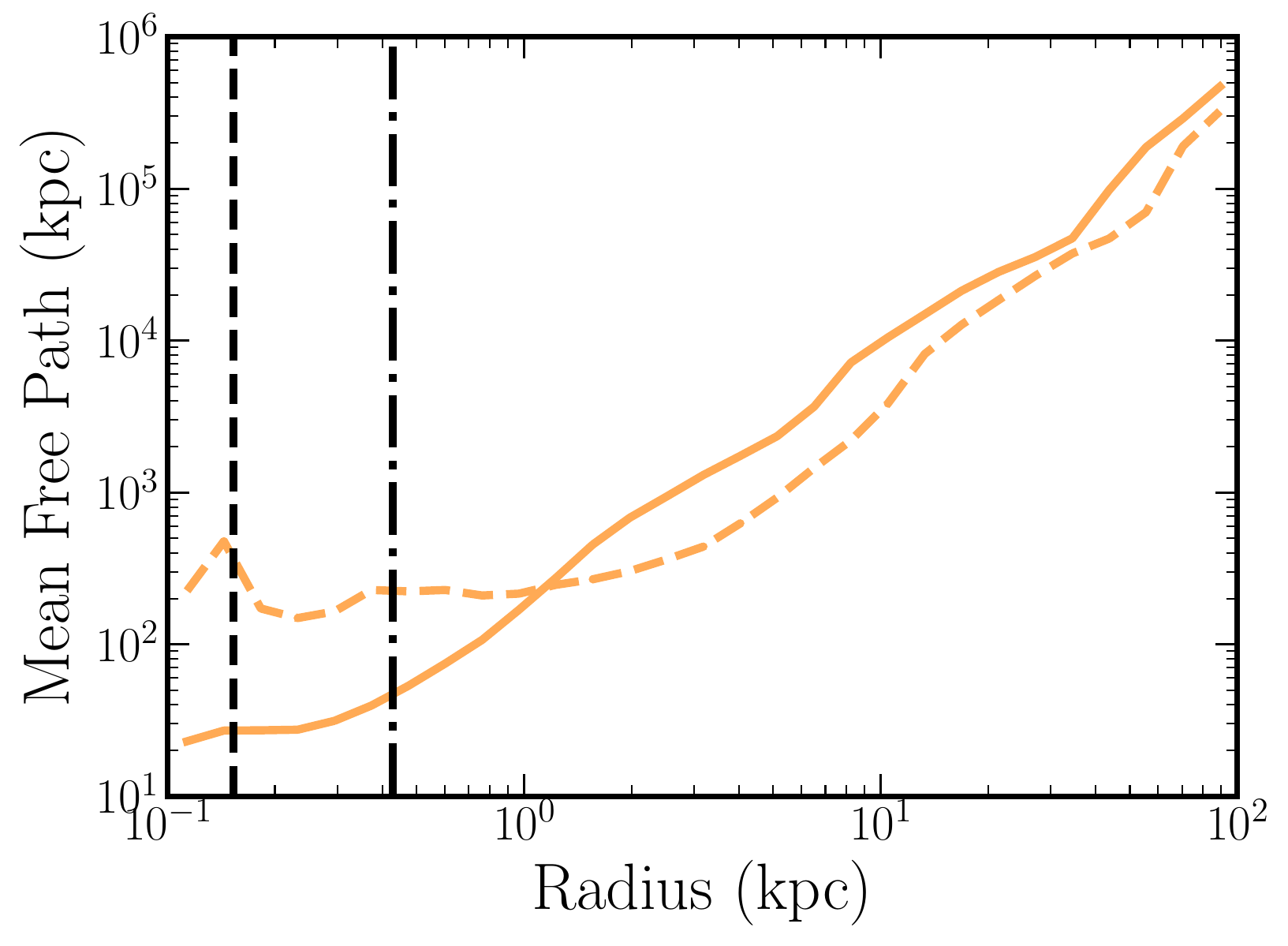}
	\includegraphics[width=.33\textwidth]{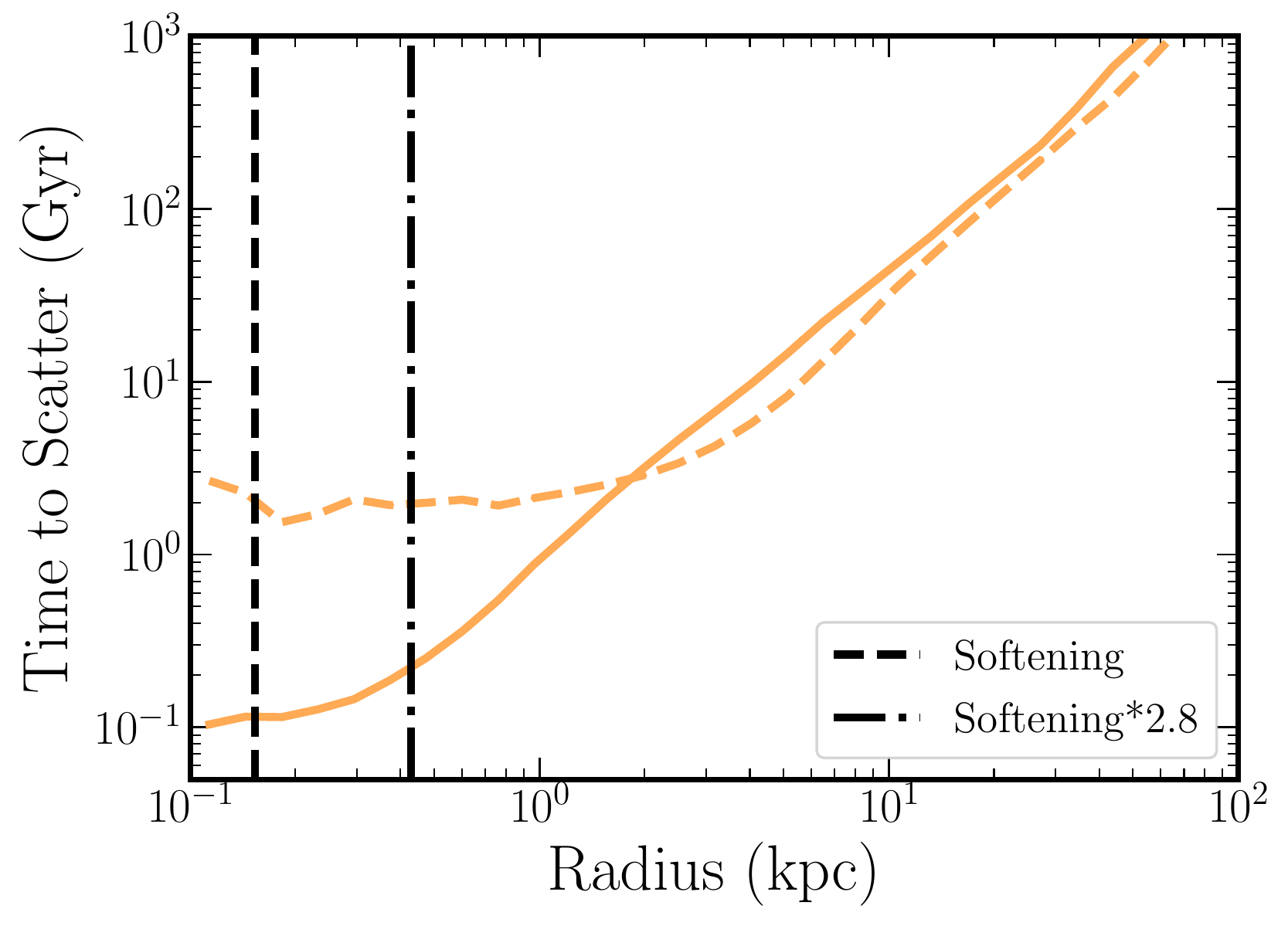}
    \caption{This figure shows the difference in the SIDM scatterings between the baryonic (solid) and DMO (dashed) SIDM simulations.
    The left panel shows the radial profiles of the average SIDM cross section calculated on a per-particle basis in logarithmic bins.
    The middle panel shows the average mean free path each DM particle sees.
    The right panel shows the average time between SIDM scatterings for each particle ("Time to Scatter").
    Each plot is a median of all particles in the same logarithmic bins.
    The DMO simulation only contains 32 particles within 300pc, so results inside this region are not well defined.
    The baryonic simulation does not suffer from this deficiency.}
    \label{fig:SIDM}
\end{figure*}

In order to investigate the effects of baryons on the relative importance of SIDM scatterings, 
Figure \ref{fig:SIDM} shows the median cross section (left), mean free path (center), and time to scatter (right). 
Because the SIDM cross sections we present are velocity dependent (see Figure \ref{fig:cross}), the average cross section profile varies with radius and the velocity dispersion profile.
Particle cross sections are calculated by taking the relative velocity of each particle with respect to each of its 32 closest neighbors and converting it to a cross section for each particle using the velocity dependent cross section. 
We then group the particles in log-spaced bins from the center of the galaxy, take the median cross section of those bins to determine the cross section radial profile. 
This profile is shown in the left panel of Figure \ref{fig:SIDM}.
We note that the DMO profile inside 300 pc is not well constrained because are only 32 particles.
The baryonic simulation does not suffer from the same information deficit as there are $\sim$1500 particles in the same region.
Outside of 30 kpc, the DMO and the baryonic SIDM simulations have similar average cross sections. The cross section profile in both simulations remains relatively constant within 10kpc. At the center of the halo (i.e. $r<10$ kpc), there is a factor $\ge$4 reduction in the average cross section in the baryonic SIDM simulation compared to the DMO simulation.

The middle panel of Figure \ref{fig:SIDM} shows the spherically averaged mean free path for the DM particles as a function of radius from the galactic center. 
This calculation is done in the same pair-wise manner as the calculation for the average cross section profile discussed in the previous paragraph. 
Here, once the cross section for each pair of particles is computed, the mean free path for that particle is calculated by multiplying its cross section by the number density of DM particles in that area.
The DMO simulation has a mean free path $\sim$1 dex larger than the baryonic simulation at the center of the halo. 
On average, the particles in the DMO simulation may interact once with another particle in the halo because the mean free path is on the order of the size of the halo. 
However, for the baryonic simulation, the mean free path at the center of the halo is much smaller than the virial radius.
Since the average cross section is smaller in the baryonic simulations, the decrease to its mean free path is due to the increase in density shown in Figure \ref{fig:DMvsB}.
Outside this region, the mean free path profiles are similar for both simulations.

The right panel of Figure \ref{fig:SIDM} calculates the current average time until a DM particle is expected to scatter with another DM particle ("Time to Scatter") as a function of radius. The Time to Scatter is calculated by dividing the mean free path of each particle by that particle's velocity in the halo's reference frame. 
The time to scatter for both simulations agree well outside 2 kpc. 
The DMO profile begins to plateau inside 3kpc, where the constant density core begins, with the particles scattering after 2-3Gyr on average. The time to scatter for the baryonic simulation decreases steadily within 3kpc, reaching one per 100 Myr ath the center, indicating that many scatterings are still occurring at redshift 0. 
Adding baryons to the simulation increases the number of scatterings in this halo by a factor of $\sim$20 compared to the DMO simulation.

\subsection{Modified DMO Simulations}
\label{sec:modified}

\begin{figure*}
	\includegraphics[width=.5\textwidth]{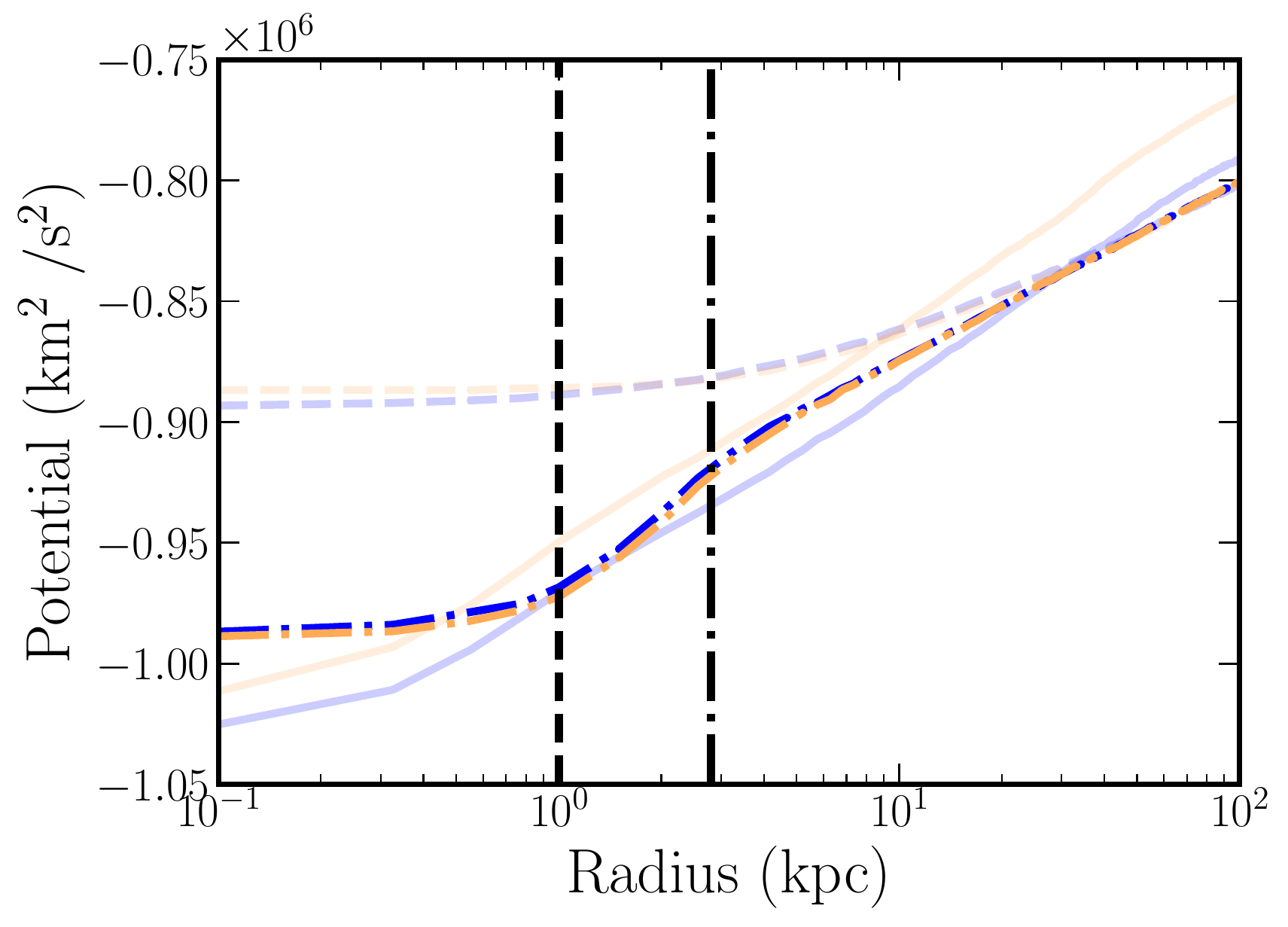}
	\includegraphics[width=.47\textwidth]{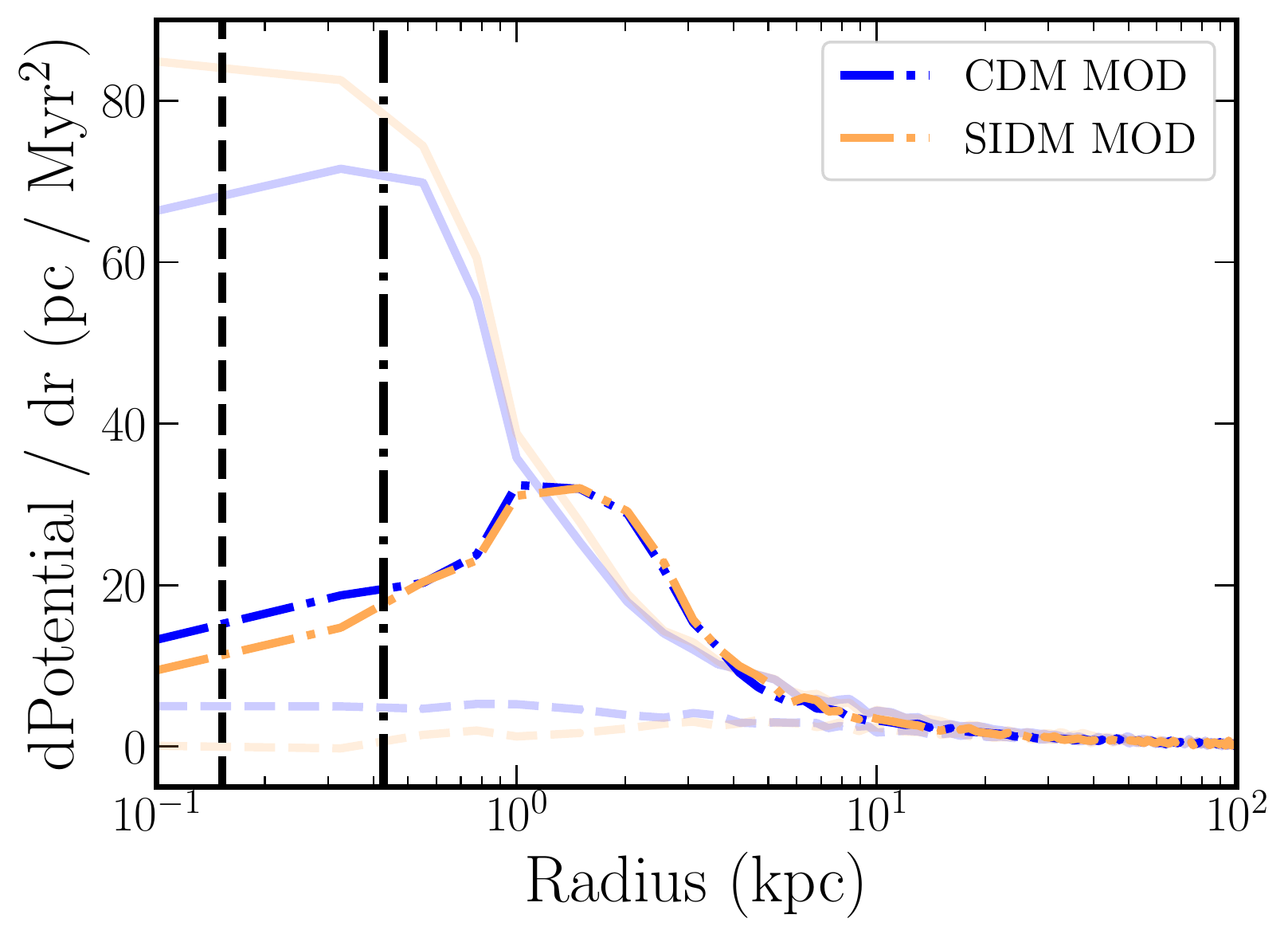}
	\includegraphics[width=.5\textwidth]{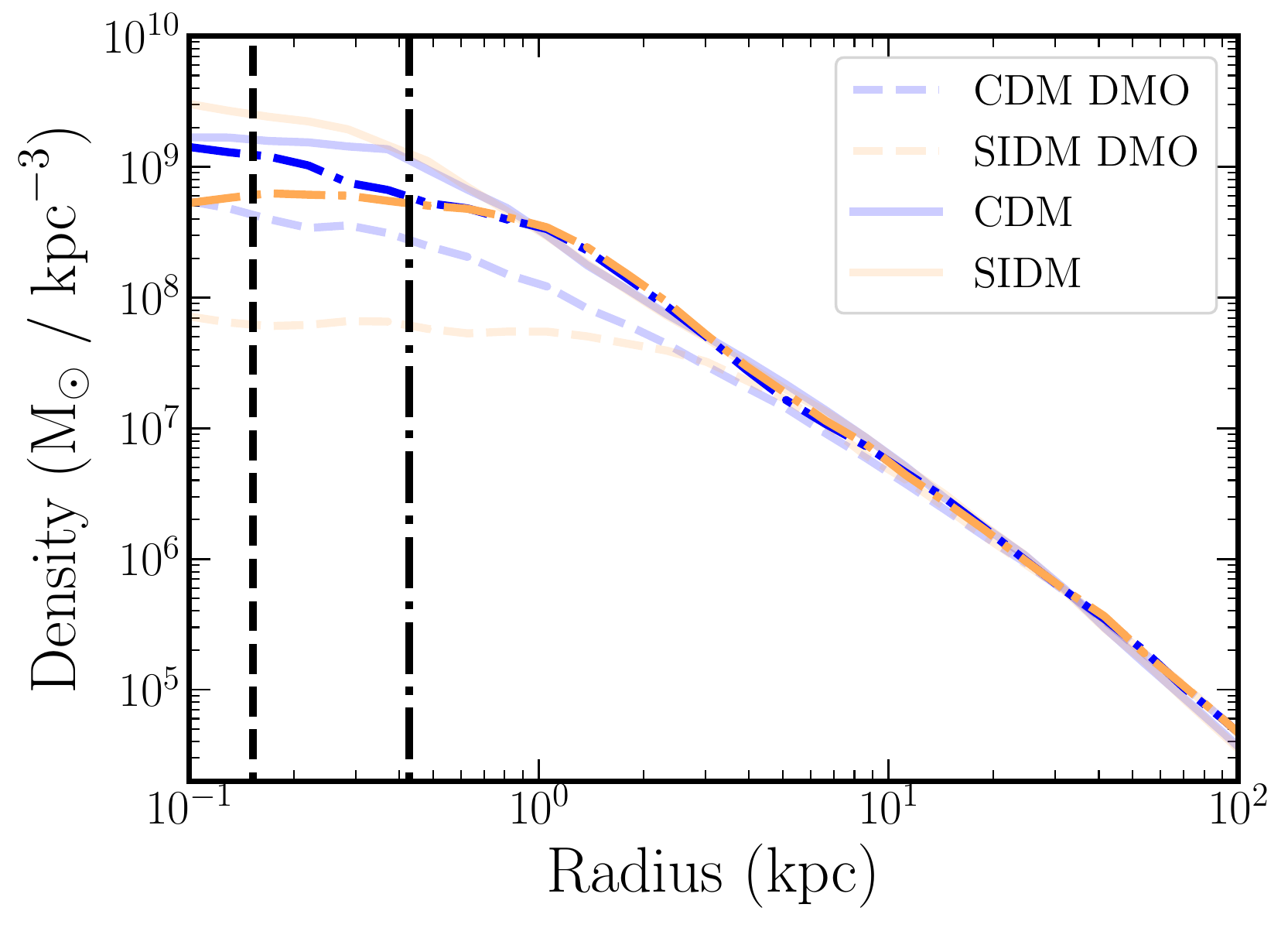}
	\includegraphics[width=.49\textwidth]{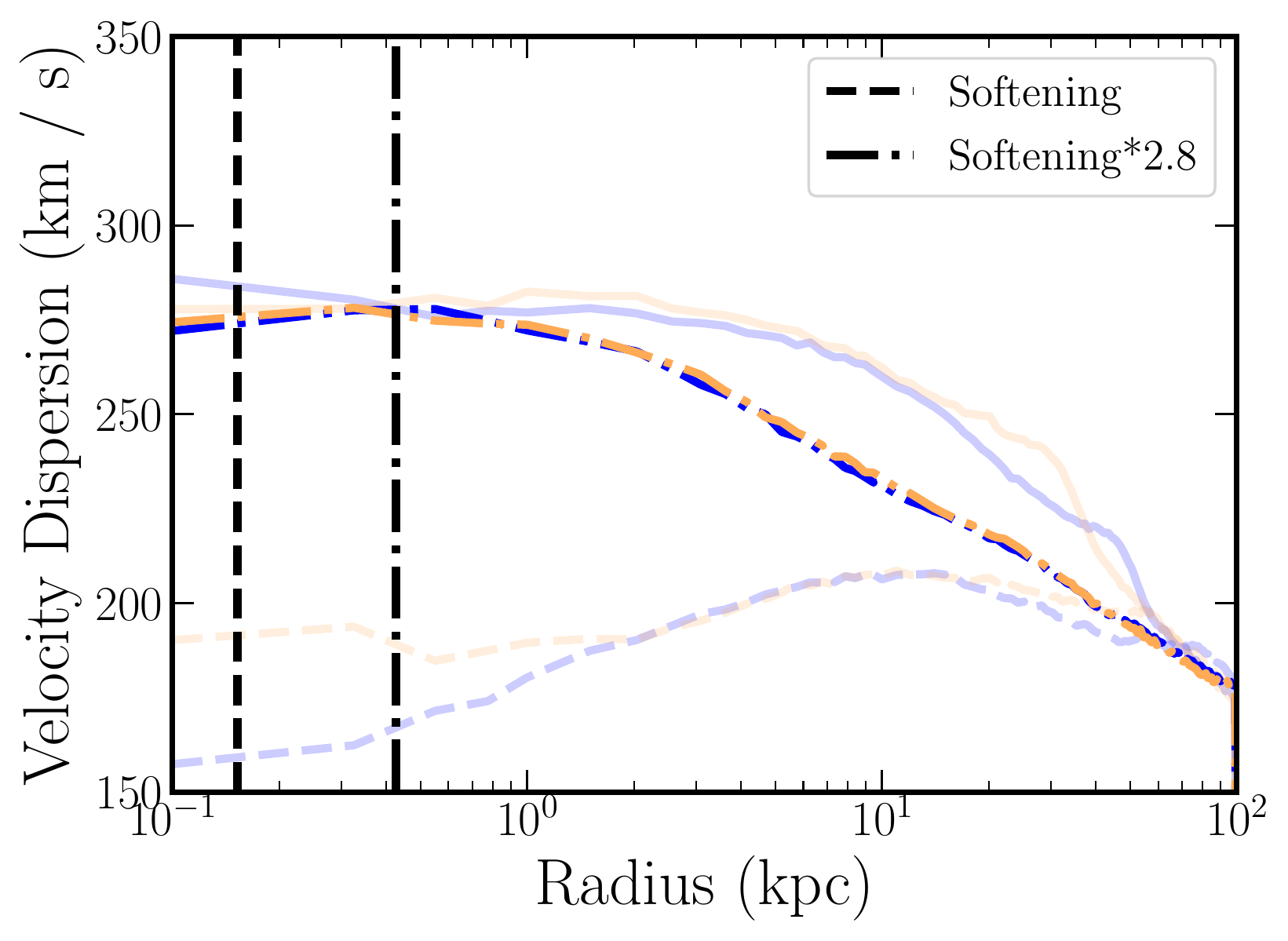}
    \caption{This figure reproduces the plots from Figure \ref{fig:DMvsB} with all six fiducial simulations. 
    The results from Figure \ref{fig:DMvsB} are repeated for reference with greater transparency. 
    The modified DMO simulations contain a particle with mass equal to the baryonic mass within 5kpc and are represented with dot-dashed lines.
    The potential profile, which is not shown in Figure \ref{fig:DMvsB}, is also included to illustrate the agreement between the modified DMO and baryonic DM halos.
    }
    \label{fig:BH}
\end{figure*}

Figure \ref{fig:BH} shows the same plots as Figure \ref{fig:DMvsB} for the two modified DMO simulations (`CDM mod' and `SIDM mod'). The results from our fiducial baryonic and DMO simulations, presented in Figure \ref{fig:DMvsB}, are reproduced here for comparison as transparent lines. 
We also provide an additional plot, the potential profile, to illustrate the changes made to the DM halo and the agreement between the baryonic and modified DMO simulations.

The top left panel shows the potential profiles for all six simulations. 
Similar to the potential derivative profiles, shown in Figure \ref{fig:DMvsB}, the profiles for the DMO and baryonic simulations deviate inside 20 kpc where the majority of baryons are present.
In this region, potential profiles of the modified DMO simulations agree well with the baryonic potential profiles.
The vertical dashed lines on this plot indicate the size of the tracer particle.
Inside 1 kpc (dashed line), the tracer has a constant density, between 1and 2.8 kpc (dot-dashed line), the tracer decreases in density away from the center of the halo.

The top right plot of Figure \ref{fig:BH} shows the derivative of the potential for the modified DMO simulations. The modified DMO simulations show an increased potential slope inside 10 kpc relative to the DMO profiles as expected from the decrease in potential shown in the top left panel. 
This increase is comparable to the fiducial baryonic simulation up to 1kpc where the tracer reaches its constant density.
Inside 1kpc, the potential slope is still increased from the DMO profiles, but not as much as the baryonic profiles.
The CDM and SIDM modified DMO simulations agree well with each other, which is to be expected since both simulation have the same massive tracer placed at the center of their halos.

The bottom left panel of Figure \ref{fig:BH} shows the spherically averaged DM density profile for the two modified DMO simulations and the results from Figure \ref{fig:DMvsB}. The results of the modified DMO simulations show an increase in the DM density from the DMO simulations inside 5 kpc.
The profiles align well with the density profile from the baryonic simulations between 5 and 1 kpc, but deviate inside 1kpc where the massive tracer has a constant density. 
The CDM and SIDM modified DMO simulations agree very well up to 2.8 $\times$ the softening length where the physics can be affected by numerical effects.

The bottom right panel of Figure \ref{fig:BH} shows the velocity dispersion profile for the two modified DMO simulations and the results from Figure \ref{fig:DMvsB}.
Again, the profile from the modified simulations deviate from the unmodified DMO simulations and are in better agreement with the full baryonic simulations, but have slightly different shapes.
Both modified DMO simulations show an isothermal core of $\sim$2 kpc.
The velocity dispersion profile in the modified DMO simulation is suppressed in the outer regions compared to the fiducial baryonic simulations. 
However, they are still increased from the DMO profiles.
Similarly to the baryonic simulations, the increase to the velocity dispersion profiles does not correspond to a decrease in the density profiles.
The SIDM and CDM modified DMO simulations agree well with each other across the entire profile.

\subsection{More SIDM cross sections}
\label{sec:const}

\begin{figure*}
	\includegraphics[width=.335\textwidth]{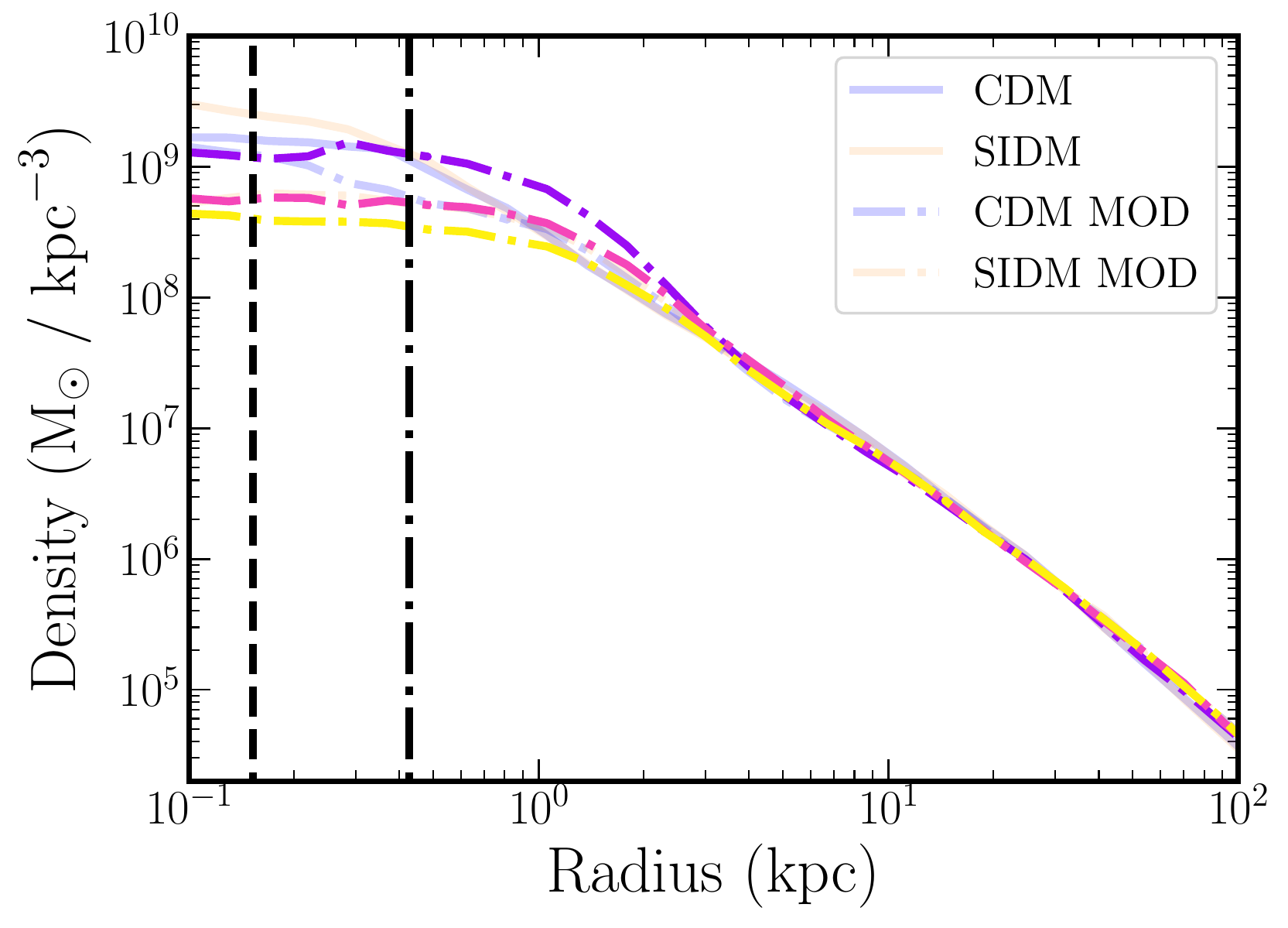}
	\includegraphics[width=.324\textwidth]{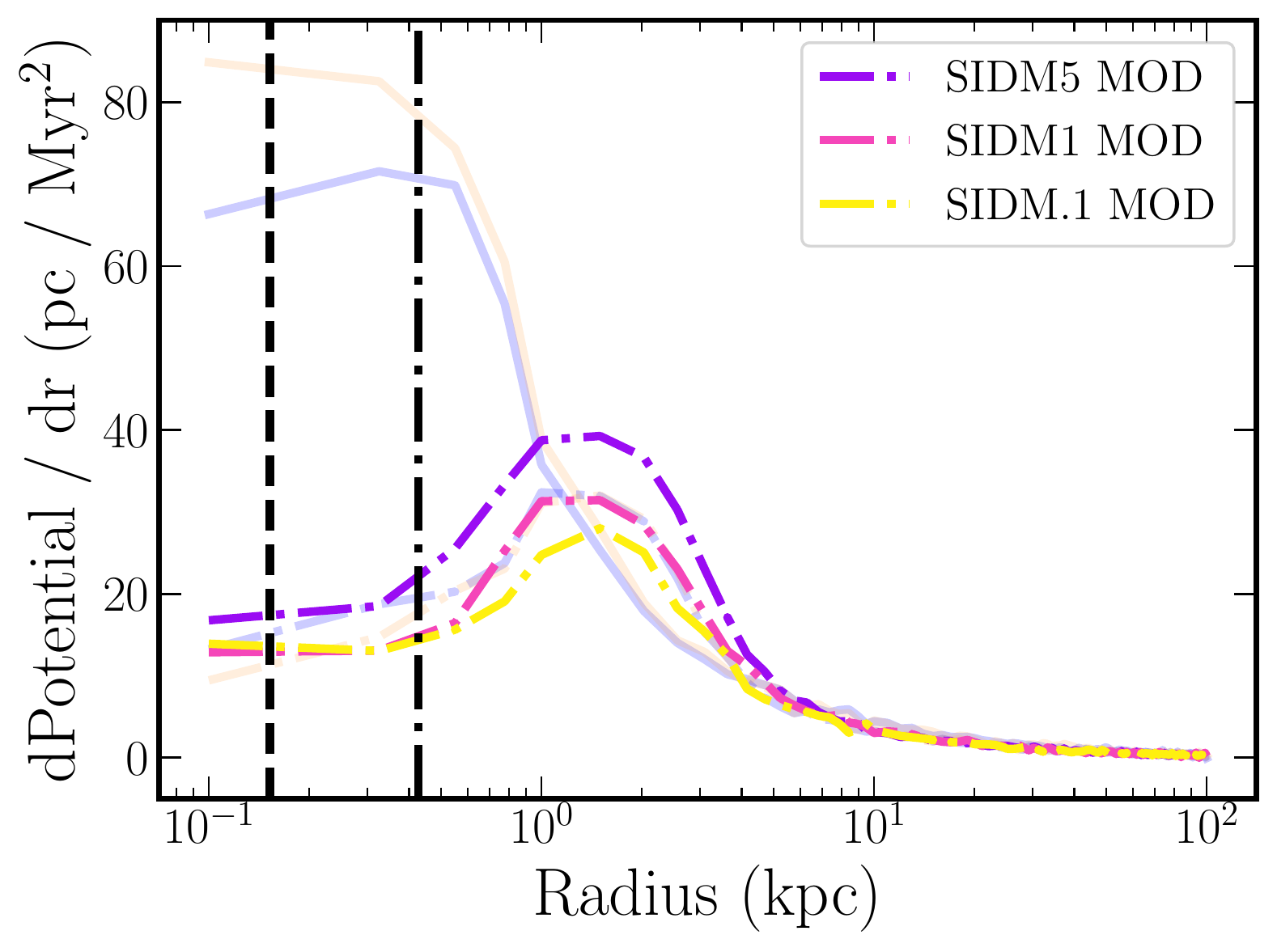}
	\includegraphics[width=.327\textwidth]{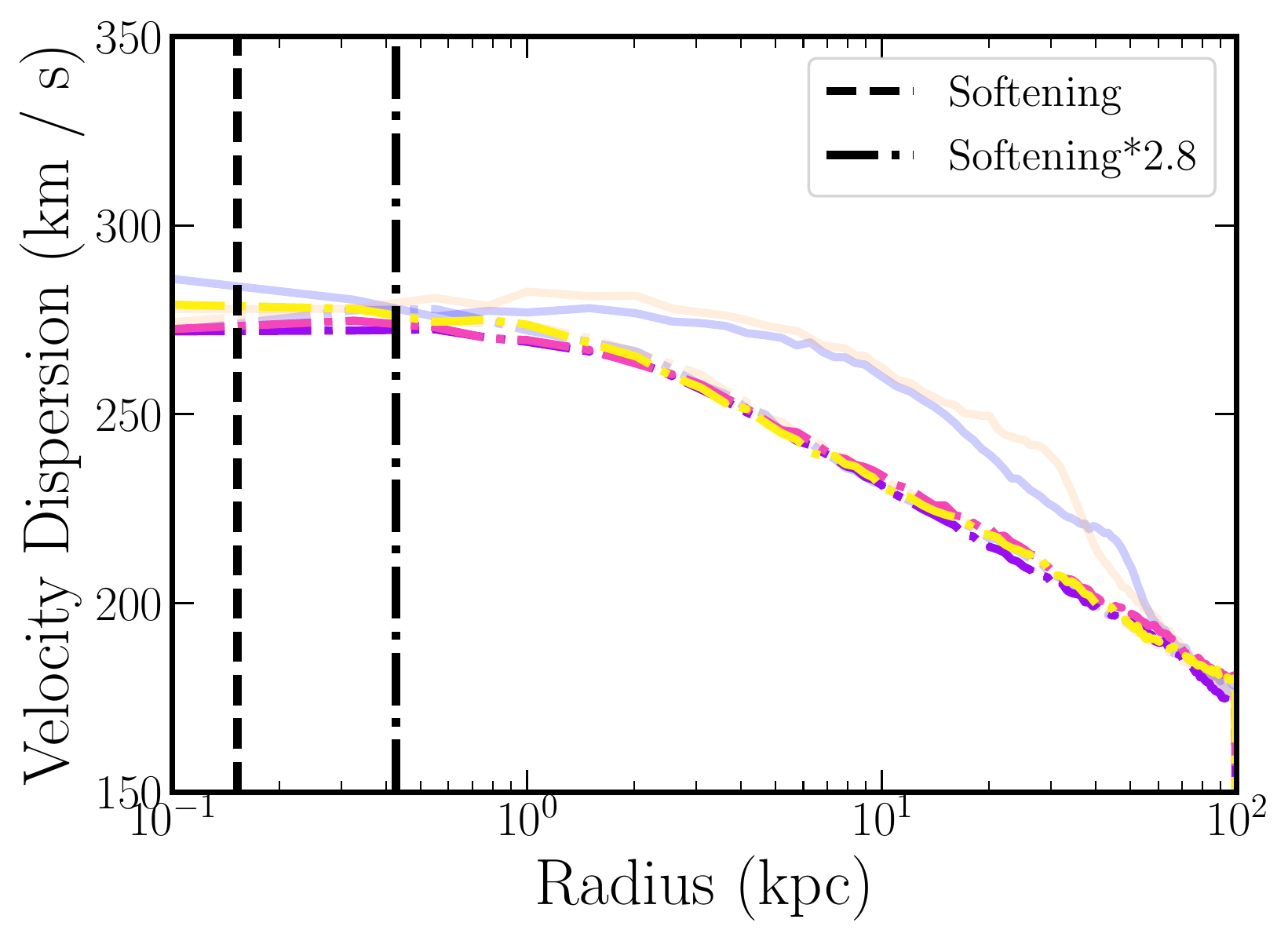}
    \caption{
    This figure reproduces the results from Figures \ref{fig:DMvsB} and \ref{fig:BH}, and introduces three additional modified DMO simulations with constant cross sections.
    The three new constant cross sections are 0.1 (yellow), 1 (pink), and 5 (purple) cm$^2$/g.
    }
    \label{fig:const}
\end{figure*}

In this section, we introduce three new modified DMO simulations. 
Each of these simulations contains an SIDM model with a constant cross section of either 5, 1, or .1 cm$^2$/g. 
The massive tracer placed at the center of these simulations is the same as those in the other modified DMO simulations.

The plots shown in Figure \ref{fig:const} are the same as Figures \ref{fig:DMvsB} and \ref{fig:BH}.
The baryonic results from Figure \ref{fig:DMvsB} and the modified DMO results from Figure \ref{fig:BH} are reproduced as transparent lines.
The profiles for the new modified DMO simulations are still represented as dot-dashed lines. 
Each cross section is represented as number next to `SIDM' in the legend (eg. `SIDM5' is a simulation with a constant cross section of 5 cm$^2$/g).

The left panel of Figure \ref{fig:const} shows the density profiles of each central halo at redshift 0.
All modified DMO simulations have the same general shape, however, SIDM1 aligns most closely with the velocity-dependent modified DMO profile.
The SIDM.1 profile shows a reduction in density in the inner few kpc and the SIDM5 profiles shown an increase in the same region.
The densities within the central 2-3 kpc increase with greater scattering cross section, indicating these halos may be in core collapse.
Outside 3kpc, the density profiles converge.

The central panel of Figure \ref{fig:const} shows the potential derivative for the three constant cross section simulations.
Again, the modified DMO simulation with constant cross section 1 cm$^2$/g aligns best with the velocity dependent modified DMO simulation.
The three constant cross section profiles also have the same general shape, but increase with increasing cross section, similar to the density profiles.

The right panel of Figure \ref{fig:const} shows the velocity dispersion profiles for the three constant cross section simulations.
Unlike the other plots in this figure, all three profiles agree nearly identically from .1 kpc to 100 kpc independent of the cross section.
Additionally, the three profiles also agree very well with the CDM and SIDM modified DMO velocity dispersion profiles from Section \ref{sec:modified}.

\subsection{Mass Dependence}
\label{sec:mass}

\begin{figure}
	\includegraphics[width=\columnwidth]{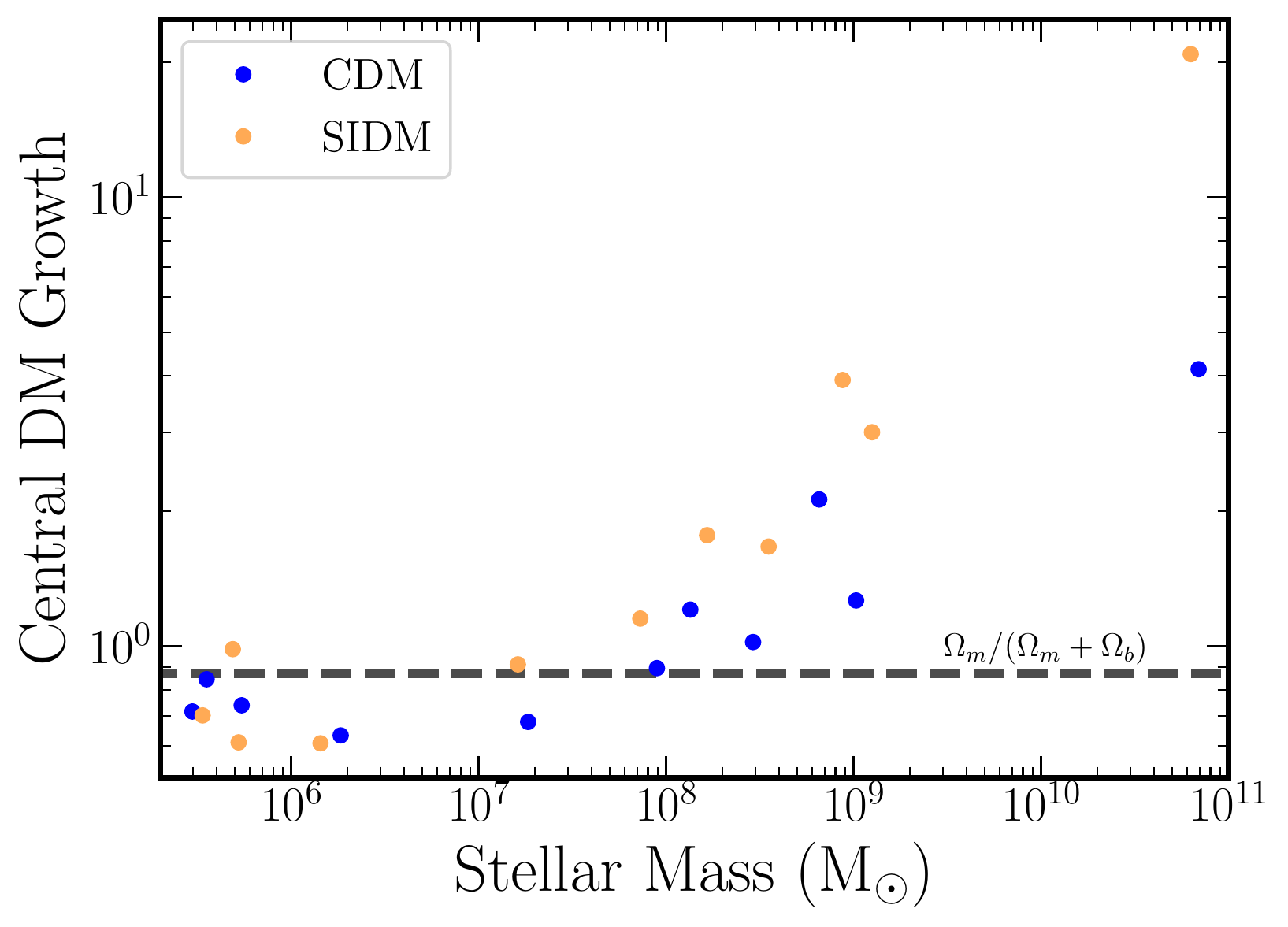}
    \caption{
    The central DM growth vs stellar mass relation for the CDM and SIDM simulations. 'Central DM Growth' refers to the ratio of DMO and baryonic density profiles  at 430 pc.
    This figure only includes galaxies that contained > 50\% of the same DM particles between the baryonic and DMO simulations.
    }
    \label{fig:mass}
\end{figure}

Figure \ref{fig:mass} shows the relative change in central DM density between the DMO and baryonic simulations as a function of stellar mass for satellites around the central halo with stellar masses greater than $10^5$ M$_\odot$.
The `Central DM Growth' is a comparison of the baryonic and DMO density profiles of the same satellite between the baryonic and DMO simulations.
We define DM growth to be the ratio of the DM density profile at 428 pc (2.8 $\times$ the softening length) in the baryonic simulations and the same region of the density profile from the DMO simulation.
A dashed line is shown at $\Omega_{m} / (\Omega_{m} + \Omega_{b})$ where there is no difference between the DMO and baryonic density profiles.
A value greater than the dashed line indicates that the density in the baryonic simulation has increased from that in the DMO simulation.
The IDs of the DM particles that make up each satellite galaxy are matched between the DMO and corresponding baryonic simulation. 
Each satellite in Figure \ref{fig:mass} is required to contain at least 50\% of the same DM particles in both simulations.
Roughly 80\% of satellites with stellar mass above $1 \times 10^5$ $M_\odot$ match this criterion.

\begin{figure}
	\includegraphics[width=\columnwidth]{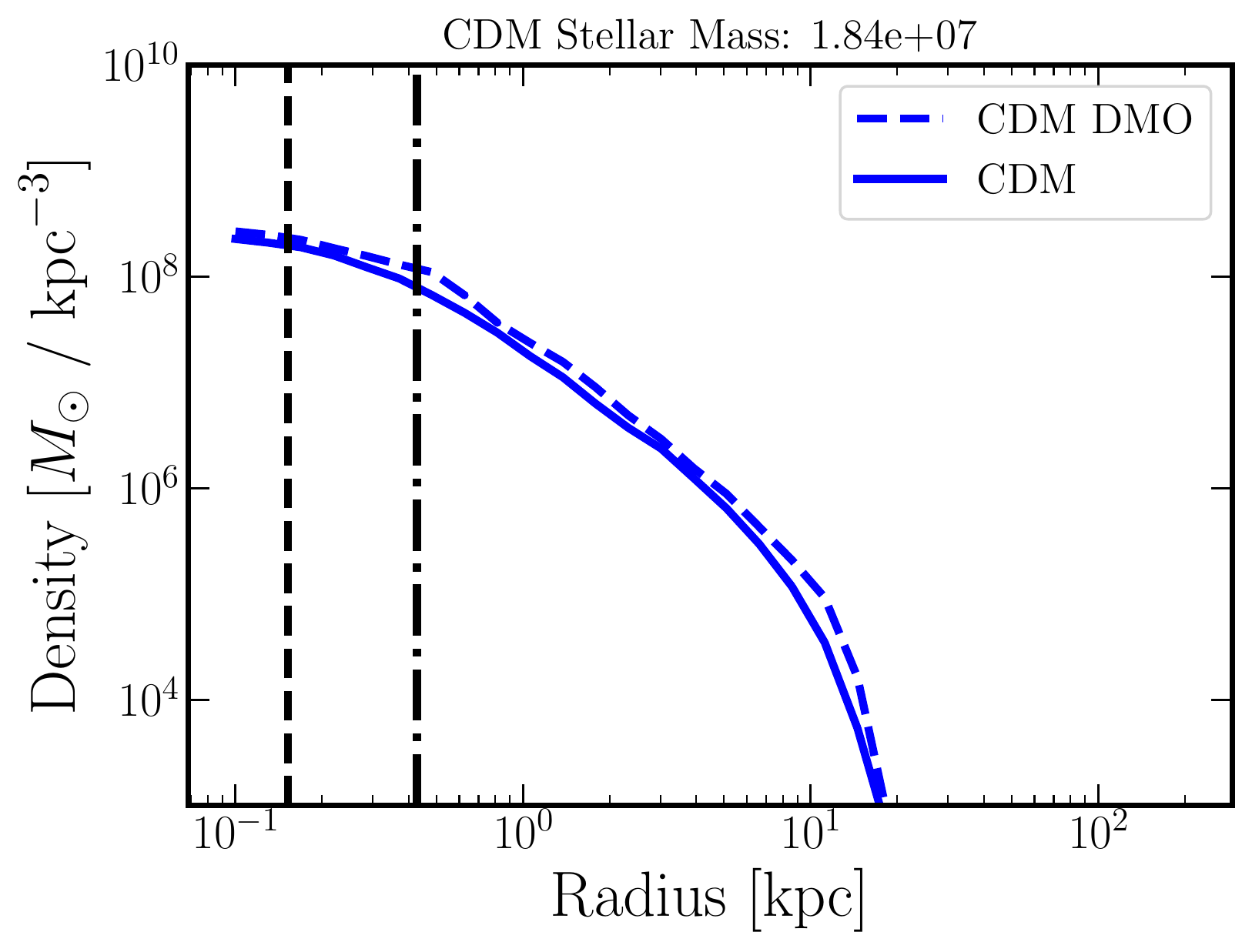}
	\includegraphics[width=\columnwidth]{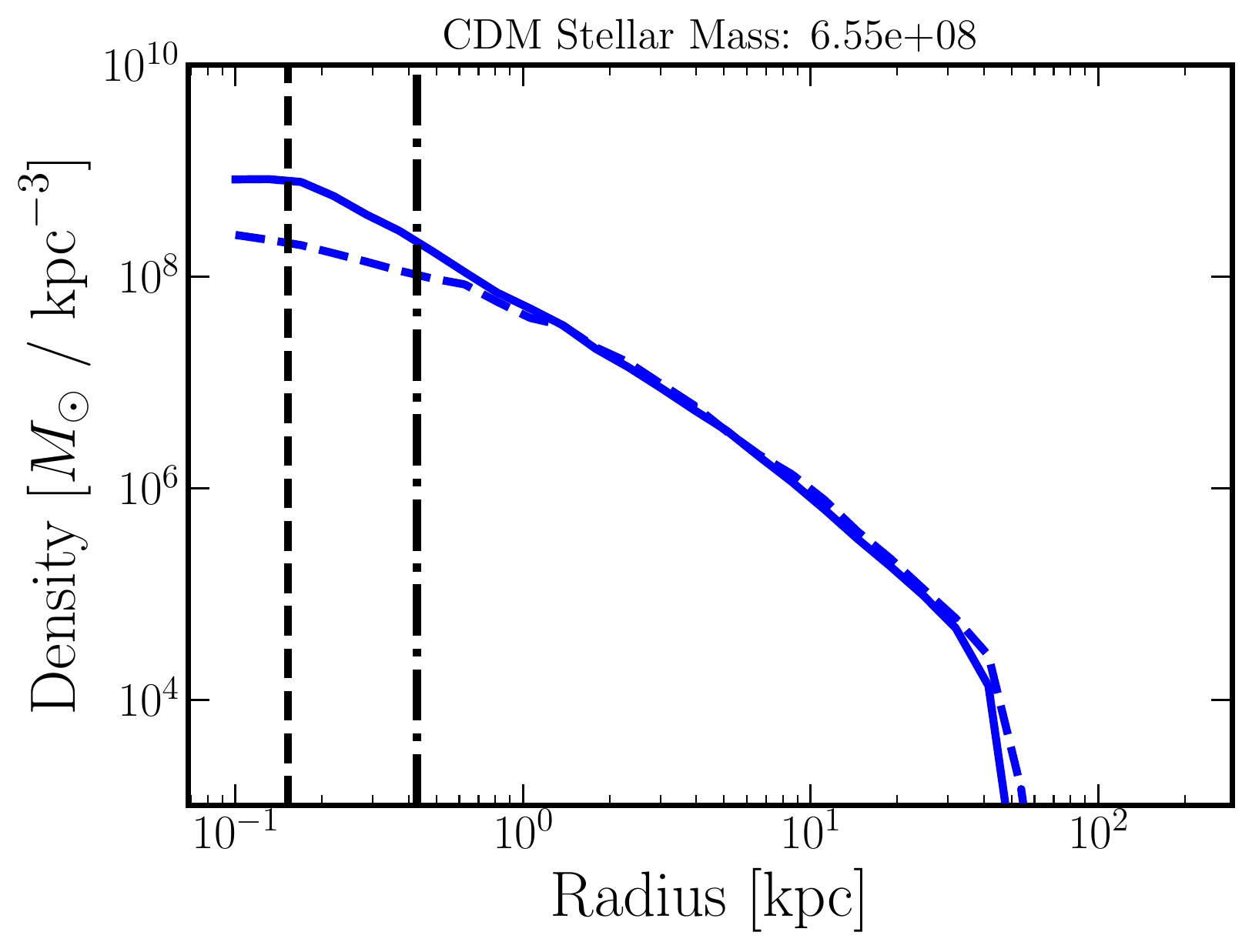}
    \caption{ This figure provides an example of the change to the DM density profile that is measured in Figure \ref{fig:mass}.
    The two satellites chosen here represent one satellite whose density profile has not undergon any changes due to the presence of baryons (top). The second satellite has enough baryonic mass to increase the density profile in the central regions (bottom).
    }
    \label{fig:mass_ind}
\end{figure}

To illustrate the change occurring in these satellites we provide two density profiles in Figure \ref{fig:mass_ind} for two sets of matched satellites in the CDM baryonic and DMO simulations.
The first satellite has a stellar mass of $6.55 \times 10^8$ M$_\odot$ and shows no deviation between the DMO and baryonic profiles.
The offset where the baryonic profile is slightly less dense than the DMO profile is due to the reduction in the matter density attributed to DM once baryons are included.
The second satellite has a stellar mass of $6.55 \times 10^8$ M$_\odot$.
Within $\sim$1-2kpc the density profiles deviate substantially as the baryonic profile becomes denser.
This deviation corresponds to a greater central DM growth depicted in Figure \ref{fig:mass} for galaxies greater than $\sim10^8$ M$_\odot$.

We find that baryons have little effect on the DM density profiles up to $10^8$ M$_\odot$, then steadily increase in strength up to the central halo at $\sim 7 \times 10^{10}$ M$_\odot$.
Up to the most massive satellites, $\sim 10^9$ M$_\odot$, there is not a large difference between the CDM and SIDM DM fraction.
At the MW mass scale there is a large (1 dex) difference between the SIDM and CDM DM fractions.
As we have seen in Figure \ref{fig:DMvsB}, the density profiles of the baryonic simulations agree very well, but the large increase in DM fraction comes from the reduced central density core from the SIDM DMO simulations.

\section{Discussion}
\label{sec:discussion}

It is apparent from Figure \ref{fig:DMvsB} that the DM density, potential slope, and velocity dispersion change drastically when baryons are included in the simulations. 
Baryons have the strongest effect in the central few kpc, but also affect these properties tens of kpc from the center, outside the central disc. 
The general shapes of these properties also change to produce qualitatively different halos than those in the DMO simulations.

The isothermal core common to SIDM halos, and the related constant density core, can be seen in the DMO SIDM profiles in Figure \ref{fig:DMvsB}.
However, we find a similar, although hotter and larger, isothermal core to the one in the DMO SIDM simulation, but no corresponding constant-density core.
This opens two questions that can be asked.

First, why do the density profiles in the baryonic simulations not contain a constant-density core when the velocity dispersion profiles contain an isothermal core? 
We see from Figure \ref{fig:DMvsB} that the potential slope, density, and velocity dispersion all increase in the central region once baryons are included.
The velocity dispersion profile specifically has increased enough to have a fundamentally different shape once baryons are included.
In the DMO simulations, the velocity dispersion profile turns over inwards of 10kpc so the SIDM scatterings allow heat to transfer from the outer regions of the halo into the central few kpc.
This increased heat forms the isothermal core present in SIDM DMO simulations and the additional pressure pushes DM out to create the constant-density core.
In the baryonic simulations, the velocity dispersion profile is greatly increased from the DMO profiles and now forms an isothermal core solely from the presence of baryons. 
At this point, there is no excess heat in the outer regions of the halo for SIDM to transfer to the center to increase support.
Without additional support, the decrease in potential from condensing baryons causes the density to increase and prevents the formation of a constant-density core.
Because of this, the role of SIDM in simulations that include baryons can be fundamentally different from its role in DMO simulations, but a detailed investigation into the heat transfer induced by SIDM can lead to an understanding of how the baryons and SIDM interplay to affect halo evolution.

Second, what is the relative impact of including baryons versus SIDM on galaxy halo structure?
As discussed in reference to Figure \ref{fig:DMvsB}, the addition of baryons increases both the velocity dispersion and density in the central region of the halo.
As a consequence of the $\sim$10kpc isothermal core that forms from baryon contraction, the SIDM scatterings no longer move heat from the outer regions of the halo inwards.
Now, DM scatterings will move heat from the hot inner region outwards.
For SIDM, the density and velocity dispersion profiles also affect the probability that two DM particles will scatter.
An increase to the velocity dispersion will increase the relative speed between the DM particles, changing the velocity dependent cross section and decreasing the number of scatterings.
An increase to the density will decrease the mean free path between two particles and increase the number of scatterings. 
Figure \ref{fig:SIDM} shows that the increase to the velocity dispersion and density profiles results in a $\sim$20$\times$ increase in the number of scatterings at the center of the halo compared to the DMO simulation.
Hence, adding baryons to an SIDM simulation increases the number of scatterings in the central region of the halo due to the increase in central density, but the increased velocity dispersion induced by baryon contraction makes it so that the scatterings will transport heat out from the center of the halo.
Since baryons maintain the isothermal core the center of the halo while SIDM transfers heat out, we find that SIDM plays a subdominant role in galaxy evolution to the effects of baryons for our SIDM model.

Figure \ref{fig:mass} shows that the results presented here do not hold equivalently for galaxies of all masses. We find that baryons have the largest affect in MW-like galaxies with diminishing effect down to a stellar mass of $\sim 10^8$ M$_\odot$. Galaxies with stellar mass less than $\sim 10^8$ M$_\odot$ are unaffected by baryon contraction and preserve their density profiles from the DMO simulations. This leaves SIDM as a promising candidate to alleviate the core-cusp and TBTF problems which reside in the $10^5$ - $10^7$ M$_\odot$ range.

We do not simulate any galaxies larger than the MW and thus cannot say how this trend will continue in larger galaxies. While these galaxies are not included in the small scale problems, they are important for SIDM theory. Galaxy mergers and clusters help to define the constraints on the SIDM cross section. Modification to the results from DMO simulations could impact the SIDM cross section constrains and could change the allowed cross sections.
\cite{2019Robertson} has simulated galaxy clusters with SIDM and baryons. They found that the inclusion of baryons changed the lensing properties of galaxy clusters and imposed different constraints on SIDM cross sections.
The baryon component in these clusters, with DM masses $10^{14}$ - $10^{15}$ M$_\odot$, is comparable to the DM mass, however more concentrated than smaller galaxies \citep{2018Lovell}.
Similarly for a MW-like galaxy, the baryon mass dominates the DM mass in the central regions and can have effects down to $10^8$ M$_\odot$.
This tells us that including baryons in SIDM simulations of nearly all galaxy sizes, especially above $10^8$ M$_\odot$, will be imperative to accurately constrain SIDM cross sections.

There are other ways in which SIDM can change the properties of the halo which are not discussed in this paper. 
For example, SIDM has been shown to make DM halos more spherical \citep{2018Tulin, 2018Sameie}.
An interesting question of how the addition of baryons into SIDM simulations will change this prediction at the MW scale remains open.
\cite{2021Vargya} investigated this question with the FIRE-2 model and found that SIDM played a subdominant role in shaping the DM and stellar halos once baryons were included.
We have also shown, in regard to the density profiles, that the effect of baryons can outweigh changes induced by SIDM.
Since both SIDM and baryons make the DM halo more spherical \citep{2019Chua, 2021Chua, 2012Vogelsberger}, a detailed understanding of the MW halo shape may allow us to distinguish between various models for both.

With the addition of the massive tracer in the modified DMO simulations, we see that the three profiles shown in Figure \ref{fig:BH} agree much more with the baryonic simulations than the DMO simulations. 
This result is generally unsurprising as the only way the DM interacts with baryons is through gravity, so adding the same amount of mass to the inner part of the halo, even if it is in a different form, produces a similar result. 
If more matter is added to the inner 10 kpc then the potential, pressure (velocity dispersion) or density will have to change in response. We see in Figure \ref{fig:BH} that all three of these profiles change dramatically once baryons are added to the simulations.

In both the CDM and SIDM modified DMO simulations, we see an increase in the potential slope, DM density, and velocity dispersion profiles from their respective DMO simulations, similar to the increase from the addition of baryons. Additionally, the differences that were present in the DMO profiles between CDM and SIDM are no longer present. 
However, the massive tracer that was added to the modified DMO simulations does not provide the same potential profile as the fully-realized implementation of baryons.
This is most apparent in the left plot of Figure \ref{fig:BH}.
The difference in the potential shape most likely accounts for the deviations in the density profile and velocity dispersion profile from the baryonic simulations.

Thus, the addition of constant cross section simulations, using the modified DMO approach presented in Section \ref{sec:const}, provide insights into how the changes to the SIDM cross section will affect MW-like halos and help disentangle halo modifications from baryons and SIDM.
As described previously, the addition of baryons increased the velocity dispersion profiles from their DMO counterparts and made the center of the halo the hottest region.
The modified DMO simulations have the same changes to the velocity dispersion profile shape as the baryonic simulations.
Therefore, it is unsurprising that larger SIDM cross sections increase the central DM density because any additional heat added to the center is transferred away and cannot add additional support.

This phenomenon, known as core collapse \citep{1968Lynden-Bell}, should also be present in the velocity-dependent cross section simulations since they share similar velocity dispersion profile shapes to the constant cross section simulations.
However, there is no difference between the CDM and SIDM density profiles presented in Figure \ref{fig:DMvsB}.
This discrepancy likely comes from the velocity dependent cross section present in our fiducial SIDM simulations.
In these simulations, as the velocity dispersion increases in the center of the halo, their cross section will decease, hindering the efficient transfer of heat out of the center of the galaxy.
\cite{2021Zeng} shows that large enough velocity dependent cross section can induce core collapse, however, this phenomenon does not appear to be present for the MW-like galaxy and velocity dependent cross section we present here.

The results and discussion presented here rely on the TNG physics model we have implemented. The TNG model does not include an explicit stellar feedback model which can repeatedly expel matter from the centers of the halos \citep{2012Pontzen}.
This formulation closely matches the implementation of the massive tracer in the modified DM simulations where the tracer sits in the center of the halo slowly accruing more mass.
This similarity most likely accounts for the agreement seen in Figure \ref{fig:BH} between the modified DMO and baryonic simulations.
If the physics model instead included bursty feedback, the potential at the center of the galaxy could drastically change over short time scales, working to push DM out from the center rather than pull it in.
This change would work to decrease the gravitational potential and increase the velocity dispersion, the opposite effect of what we see with the TNG model.

\cite{2021Sameie} has presented the only other suite of simulations which focus their analysis on MW-like galaxies with SIDM and baryons.
Their simulations follow the FIRE-2 model \citep{2014Hopkins} and differ from ours in that they include an explicit feedback model and constant cross sections for their SIDM.
They found that the difference between the DM density profiles in DMO and baryonic simulations was larger in SIDM simulations than CDM simulations.
They reasoned that this larger increase in SIDM DM profiles was caused by baryon contraction causing the SIDM halos to transition faster from core-expansion to core-contraction.
From this, they concluded that SIDM halos have a stronger reaction to the presence of baryons than CDM halos.

This result agrees with our findings for our constant cross section simulations (see Section \ref{sec:const}) where the central DM density increases with SIDM cross section.
On face value, Figure \ref{fig:mass} of this paper also agrees with this result from \cite{2021Sameie} as the SIDM halos show a larger central DM growth than CDM halos at high stellar masses when comparing DMO to baryonic simulations.
However, as we discussed earlier, the DM density profiles for the two central halos in the baryonic simulations are very similar, even though the SIDM profile has a larger change from its DMO profile (see Figure \ref{fig:DMvsB}).
The apparent increase in response to the presence of baryons in our fiducial SIDM simulations originates from differences in the corresponding DMO simulations rather than differences in the resulting baryonic simulations.
Specifically, while the DM density profiles in the baryonic simulations are nearly identical, the density profile in the SIDM DMO simulation is nearly an order of magnitude less dense at the center of the halo due to the constant density core.  
Because Figure \ref{fig:mass} shows the increase in the DM density from DMO to baryonic simulations, the order of magnitude difference in density in the DMO simulations results in an order of magintue increase in the Central DM Growth when comparing CDM and SIDM simulations.
This increase does not come from a difference in the baryonic simulations as their density profiles agree very well.
Therefore, we do not see that our fiducial SIDM halos are more responsive to the presence of baryons than the CDM halos.
Instead, we find that comparing the effects of adding baryons to a DMO simulation gives one conclusion, but comparing baryonic SIDM simulations to CDM simulations provides another.
From this we emphasize that comparing changes from DMO to baryonic simulations alone may not provide a clear benchmark to understand how baryons and SIDM interact during halo formation and evolution.

The main difference between our fiducial simulations, where we do not see a different response to baryons in CDM and SIDM halos, and our constant cross section simulations, where we do see a difference in responsiveness, is the velocity-dependent nature of the SIDM cross sections.
Given that the DM density profiles in our fiducial simulations do not show a different reaction to the presence of baryons in SIDM and CDM, we find that the claim that SIDM halos have a stronger reaction to the presence of baryons does not hold for all SIDM models.
Instead, we find that a detailed view of the heat transfer within a galaxy provides a more robust explanation for the interaction between baryons and SIDM.

As discussed earlier, we find two main factors that contribute to the DM density profile: SIDM scatterings which transport heat either toward or away from the center and baryon contraction which pulls DM toward the center of the halo.
In CDM, baryon contraction becomes less efficient as the DM in the center of the halo heats up and supplies support to resist additional DM being transported to the center.
Once SIDM is introduced, this heat can be transported out from the center of the halo, removing the support, and allowing DM to continue to fall into the center of the halo.
This mechanism can cause the SIDM DM profiles to be much steeper than the CDM profiles, as seen in \cite{2021Sameie} and Section \ref{sec:const} of this paper.
However, this transport becomes less efficient in models with a velocity dependent cross section because the heat at the center of the halo increases the relative velocity between particles and decreases the number of SIDM scatterings.

Explicit feedback models, like the one used in \cite{2021Sameie}, have a third factor that contributes to the DM density profile: stronger stellar feedback.
The stronger feedback can push DM out from the center of the halo, counteracting the effects of baryon contraction which can from a constant-density core.
This additional feature complicates the picture outlined here, and since our models do not contain an explicit feedback model, we do not explore the interplay between explicit stellar feedback, baryon contraction, and heat transfer through SIDM scatterings.

This analysis is limited to the one velocity dependent and three constant elastic cross sections presented in Sections \ref{sec:methods} and \ref{sec:const} respectively.
Various other SIDM models have been tested at the MW scale and shown to produce different results from those shown here \citep{2016Vogelsberger, 2021Zeng, 2021Sameie}.
Models with elastic collisions will likely behave similarly to the models we present here, but with variation on a halo-to-halo basis as seen in \cite{2021Sameie}.
\cite{2021ChuaA} simulated a MW-like galaxy with inelastic collisions and found that their SIDM models produced larger cores in their galaxy.
Including an inelastic SIDM model in a hydrodynamic simulation may provide additional insight into how baryons and SIDM can interact during galaxy evolution.

\section{Conclusions}
\label{sec:conclustion}

We presented the first results from simulating a MW-like galaxy with IllustrisTNG physics and SIDM to better understand how the presence of baryons affect SIDM predictions. We also presented the results from modified DMO simulations to help disentangle the effects baryons and SIDM have on their host halo. Our main conclusions are as follows:

\begin{enumerate}
    \item Including both SIDM and IllustrisTNG physics produces realistic MW-like galaxies. These models create disc-dominated galaxies that match a host of observations, such as the Tully Fisher relation, stellar mass - halo mass relation, and size - mass relation (see Section \ref{sec:realGal}).
    \item Baryon contraction is the main mechanism which increases the DM density, potential slope, and velocity dispersion profiles in both CDM and SIDM simulations (see Section \ref{sec:bvsDMO}).
    \item Baryon contraction increases the number of SIDM scatterings by a factor of 40 in the inner 1kpc compared to the same galaxy without baryons (see Section \ref{sec:sidm}).
    \item The effect of baryons following the IllustrisTNG framework on the DM halo can be reasonably distilled to a deeper potential. Simplified simulations which only include DM and a massive tracer particle produce similar DM halos to a fully realized hydrodynamical simulation for both CDM and SIDM (see Section \ref{sec:modified}).
    \item MW-like galaxies with constant cross section SIDM are more likely to undergo core collapse than velocity dependent cross sections once baryons are included. The increased central densities caused by condensing baryons decrease the mean inter-particle separation which cause more scatterings in constant cross section models. However, this effect is mitigated by an increased velocity dispersion which lowers the cross section for our velocity-dependent models (see Section \ref{sec:const}).
    \item The effects of baryon contraction are strongest for MW-like galaxies and decrease until $\sim 10^8$ M$_\odot$ where there is a negligible change from DMO simulations (see Section \ref{sec:mass}).
\end{enumerate}

We find that the presence of baryons can strongly affect the predictions from DMO simulations for galaxies with stellar mass above $\sim 10^8$ M$_\odot$.
At the MW-mass scale, galaxies can form an isothermal core solely from the presence of baryons.
This fundamentally changes the nature of SIDM as it now transfers heat away from the center of the galaxy.
However, we find that the role of SIDM in MW-like galaxies is subdominant to the effects of baryons in dictating halo properties.
The baryonic effects abate down to galaxies with a stellar mass less than $10^8$ M$_\odot$ where the results from baryonic and DMO simulations agree.
The small-scale problems present in CDM simulations, namely the core-cusp and TBTF problems, lay below this threshold ($\lesssim10^7$ M$_\odot$).
This leaves SIDM as a strong contender to alleviate these problems.
However, above the $\sim 10^8$ M$_\odot$ threshold, baryons must be included to accurately make predictions with SIDM.
This could change the currently allowed cross-sections for SIDM that form the constraints from galaxy cluster simulations.

\section*{Acknowledgements}

The authors acknowledge helpful conversation with Mikhail Medvedev, Ryan Low, and Rakshak Adhikari. 
The authors acknowledge University of Florida Research Computing for providing computational resources and support that have contributed to the research results reported in this publication.
JCR acknowledges support from the University of Florida Graduate School’s Graduate Research Fellowship.
PT acknowledges support from NSF grant AST-1909933, AST-2008490, and NASA ATP Grant 80NSSC20K0502.

MV acknowledges support through NASA ATP 19-ATP19-0019, 19-ATP19-0020, 19-ATP19-0167, and NSF grants AST-1814053, AST-1814259, AST-1909831, AST-2007355 and AST-2107724.

\section*{Data Availability}

The data and code used to produce this paper can be made available upon reasonable request to the corresponding author.

\bibliographystyle{mnras}
\bibliography{citations}

\begin{thebibliography}{}
\makeatletter
\relax
\def\mn@urlcharsother{\let\do\@makeother \do\$\do\&\do\#\do\^\do\_\do\%\do\~}
\def\mn@doi{\begingroup\mn@urlcharsother \@ifnextchar [ {\mn@doi@}
  {\mn@doi@[]}}
\def\mn@doi@[#1]#2{\def\@tempa{#1}\ifx\@tempa\@empty \href
  {http://dx.doi.org/#2} {doi:#2}\else \href {http://dx.doi.org/#2} {#1}\fi
  \endgroup}
\def\mn@eprint#1#2{\mn@eprint@#1:#2::\@nil}
\def\mn@eprint@arXiv#1{\href {http://arxiv.org/abs/#1} {{\tt arXiv:#1}}}
\def\mn@eprint@dblp#1{\href {http://dblp.uni-trier.de/rec/bibtex/#1.xml}
  {dblp:#1}}
\def\mn@eprint@#1:#2:#3:#4\@nil{\def\@tempa {#1}\def\@tempb {#2}\def\@tempc
  {#3}\ifx \@tempc \@empty \let \@tempc \@tempb \let \@tempb \@tempa \fi \ifx
  \@tempb \@empty \def\@tempb {arXiv}\fi \@ifundefined
  {mn@eprint@\@tempb}{\@tempb:\@tempc}{\expandafter \expandafter \csname
  mn@eprint@\@tempb\endcsname \expandafter{\@tempc}}}

\bibitem[\protect\citeauthoryear{{Blumenthal}, {Faber}, {Flores}  \&
  {Primack}}{{Blumenthal} et~al.}{1986}]{1986Blumenthal}
{Blumenthal} G.~R.,  {Faber} S.~M.,  {Flores} R.,   {Primack} J.~R.,  1986,
  \mn@doi [\apj] {10.1086/163867}, \href
  {https://ui.adsabs.harvard.edu/abs/1986ApJ...301...27B} {301, 27}

\bibitem[\protect\citeauthoryear{{Boveia} \& {Doglioni}}{{Boveia} \&
  {Doglioni}}{2018}]{2018Boveia}
{Boveia} A.,  {Doglioni} C.,  2018, \mn@doi [Annual Review of Nuclear and
  Particle Science] {10.1146/annurev-nucl-101917-021008}, \href
  {https://ui.adsabs.harvard.edu/abs/2018ARNPS..68..429B} {68, 429}

\bibitem[\protect\citeauthoryear{{Boylan-Kolchin}, {Bullock}  \&
  {Kaplinghat}}{{Boylan-Kolchin} et~al.}{2011}]{2011Boylan}
{Boylan-Kolchin} M.,  {Bullock} J.~S.,   {Kaplinghat} M.,  2011, \mn@doi
  [\mnras] {10.1111/j.1745-3933.2011.01074.x}, \href
  {https://ui.adsabs.harvard.edu/abs/2011MNRAS.415L..40B} {415, L40}

\bibitem[\protect\citeauthoryear{{Brooks} \& {Zolotov}}{{Brooks} \&
  {Zolotov}}{2014}]{2014Brooks}
{Brooks} A.~M.,  {Zolotov} A.,  2014, \mn@doi [\apj]
  {10.1088/0004-637X/786/2/87}, \href
  {https://ui.adsabs.harvard.edu/abs/2014ApJ...786...87B} {786, 87}

\bibitem[\protect\citeauthoryear{{Bullock} \& {Boylan-Kolchin}}{{Bullock} \&
  {Boylan-Kolchin}}{2017}]{2017Bullock}
{Bullock} J.~S.,  {Boylan-Kolchin} M.,  2017, \mn@doi [\araa]
  {10.1146/annurev-astro-091916-055313}, \href
  {https://ui.adsabs.harvard.edu/abs/2017ARA&A..55..343B} {55, 343}

\bibitem[\protect\citeauthoryear{{Callingham}, {Cautun}, {Deason}, {Frenk},
  {Grand}, {Marinacci}  \& {Pakmor}}{{Callingham}
  et~al.}{2020}]{2020Callingham}
{Callingham} T.~M.,  {Cautun} M.,  {Deason} A.~J.,  {Frenk} C.~S.,  {Grand} R.
  J.~J.,  {Marinacci} F.,   {Pakmor} R.,  2020, \mn@doi [\mnras]
  {10.1093/mnras/staa1089}, \href
  {https://ui.adsabs.harvard.edu/abs/2020MNRAS.495...12C} {495, 12}

\bibitem[\protect\citeauthoryear{{Carlson}, {Machacek}  \& {Hall}}{{Carlson}
  et~al.}{1992}]{1992Machacek}
{Carlson} E.~D.,  {Machacek} M.~E.,   {Hall} L.~J.,  1992, \mn@doi [\apj]
  {10.1086/171833}, \href
  {https://ui.adsabs.harvard.edu/abs/1992ApJ...398...43C} {398, 43}

\bibitem[\protect\citeauthoryear{{Carton Zeng}, {Peter}, {Du}, {Benson}, {Kim},
  {Jiang}, {Cyr-Racine}  \& {Vogelsberger}}{{Carton Zeng}
  et~al.}{2021}]{2021Zeng}
{Carton Zeng} Z.,  {Peter} A. H.~G.,  {Du} X.,  {Benson} A.,  {Kim} S.,
  {Jiang} F.,  {Cyr-Racine} F.-Y.,   {Vogelsberger} M.,  2021, arXiv e-prints,
  \href {https://ui.adsabs.harvard.edu/abs/2021arXiv211000259C} {p.
  arXiv:2110.00259}

\bibitem[\protect\citeauthoryear{{Chan}, {Kere{\v{s}}}, {O{\~n}orbe},
  {Hopkins}, {Muratov}, {Faucher-Gigu{\`e}re}  \& {Quataert}}{{Chan}
  et~al.}{2015}]{2015Chan}
{Chan} T.~K.,  {Kere{\v{s}}} D.,  {O{\~n}orbe} J.,  {Hopkins} P.~F.,  {Muratov}
  A.~L.,  {Faucher-Gigu{\`e}re} C.~A.,   {Quataert} E.,  2015, \mn@doi [\mnras]
  {10.1093/mnras/stv2165}, \href
  {https://ui.adsabs.harvard.edu/abs/2015MNRAS.454.2981C} {454, 2981}

\bibitem[\protect\citeauthoryear{{Chua}, {Pillepich}, {Vogelsberger}  \&
  {Hernquist}}{{Chua} et~al.}{2019}]{2019Chua}
{Chua} K. T.~E.,  {Pillepich} A.,  {Vogelsberger} M.,   {Hernquist} L.,  2019,
  \mn@doi [\mnras] {10.1093/mnras/sty3531}, \href
  {https://ui.adsabs.harvard.edu/abs/2019MNRAS.484..476C} {484, 476}

\bibitem[\protect\citeauthoryear{{Chua}, {Vogelsberger}, {Pillepich}  \&
  {Hernquist}}{{Chua} et~al.}{2021a}]{2021Chua}
{Chua} K. T.~E.,  {Vogelsberger} M.,  {Pillepich} A.,   {Hernquist} L.,  2021a,
  arXiv e-prints, \href {https://ui.adsabs.harvard.edu/abs/2021arXiv210900012C}
  {p. arXiv:2109.00012}

\bibitem[\protect\citeauthoryear{{Chua}, {Dibert}, {Vogelsberger}  \&
  {Zavala}}{{Chua} et~al.}{2021b}]{2021ChuaA}
{Chua} K. T.~E.,  {Dibert} K.,  {Vogelsberger} M.,   {Zavala} J.,  2021b,
  \mn@doi [\mnras] {10.1093/mnras/staa3315}, \href
  {https://ui.adsabs.harvard.edu/abs/2021MNRAS.500.1531C} {500, 1531}

\bibitem[\protect\citeauthoryear{{Clowe}, {Brada{\v{c}}}, {Gonzalez},
  {Markevitch}, {Randall}, {Jones}  \& {Zaritsky}}{{Clowe}
  et~al.}{2006}]{2006Clowe}
{Clowe} D.,  {Brada{\v{c}}} M.,  {Gonzalez} A.~H.,  {Markevitch} M.,  {Randall}
  S.~W.,  {Jones} C.,   {Zaritsky} D.,  2006, \mn@doi [\apjl] {10.1086/508162},
  \href {https://ui.adsabs.harvard.edu/abs/2006ApJ...648L.109C} {648, L109}

\bibitem[\protect\citeauthoryear{{Cyr-Racine}, {Sigurdson}, {Zavala},
  {Bringmann}, {Vogelsberger}  \& {Pfrommer}}{{Cyr-Racine}
  et~al.}{2016}]{2016Cyr-Racine}
{Cyr-Racine} F.-Y.,  {Sigurdson} K.,  {Zavala} J.,  {Bringmann} T.,
  {Vogelsberger} M.,   {Pfrommer} C.,  2016, \mn@doi [\prd]
  {10.1103/PhysRevD.93.123527}, \href
  {https://ui.adsabs.harvard.edu/abs/2016PhRvD..93l3527C} {93, 123527}

\bibitem[\protect\citeauthoryear{{Engler} et~al.,}{{Engler}
  et~al.}{2021}]{2021Engler}
{Engler} C.,  et~al., 2021, \mn@doi [\mnras] {10.1093/mnras/stab2437}, \href
  {https://ui.adsabs.harvard.edu/abs/2021MNRAS.507.4211E} {507, 4211}

\bibitem[\protect\citeauthoryear{{Fitts} et~al.,}{{Fitts}
  et~al.}{2017}]{2017Fitts}
{Fitts} A.,  et~al., 2017, \mn@doi [\mnras] {10.1093/mnras/stx1757}, \href
  {https://ui.adsabs.harvard.edu/abs/2017MNRAS.471.3547F} {471, 3547}

\bibitem[\protect\citeauthoryear{{Fitts} et~al.,}{{Fitts}
  et~al.}{2019}]{2019Fitts}
{Fitts} A.,  et~al., 2019, \mn@doi [\mnras] {10.1093/mnras/stz2613}, \href
  {https://ui.adsabs.harvard.edu/abs/2019MNRAS.490..962F} {490, 962}

\bibitem[\protect\citeauthoryear{{Fry} et~al.,}{{Fry} et~al.}{2015}]{2015Fry}
{Fry} A.~B.,  et~al., 2015, \mn@doi [\mnras] {10.1093/mnras/stv1330}, \href
  {https://ui.adsabs.harvard.edu/abs/2015MNRAS.452.1468F} {452, 1468}

\bibitem[\protect\citeauthoryear{{Gadotti}}{{Gadotti}}{2009}]{2009Gadotti}
{Gadotti} D.~A.,  2009, \mn@doi [\mnras] {10.1111/j.1365-2966.2008.14257.x},
  \href {https://ui.adsabs.harvard.edu/abs/2009MNRAS.393.1531G} {393, 1531}

\bibitem[\protect\citeauthoryear{{Grand} et~al.,}{{Grand}
  et~al.}{2017}]{2017Grand}
{Grand} R. J.~J.,  et~al., 2017, \mn@doi [\mnras] {10.1093/mnras/stx071}, \href
  {https://ui.adsabs.harvard.edu/abs/2017MNRAS.467..179G} {467, 179}

\bibitem[\protect\citeauthoryear{{Hahn} \& {Abel}}{{Hahn} \&
  {Abel}}{2011}]{2011Hahn}
{Hahn} O.,  {Abel} T.,  2011, \mn@doi [\mnras]
  {10.1111/j.1365-2966.2011.18820.x}, \href
  {https://ui.adsabs.harvard.edu/abs/2011MNRAS.415.2101H} {415, 2101}

\bibitem[\protect\citeauthoryear{{Hopkins}, {Kere{\v{s}}}, {O{\~n}orbe},
  {Faucher-Gigu{\`e}re}, {Quataert}, {Murray}  \& {Bullock}}{{Hopkins}
  et~al.}{2014}]{2014Hopkins}
{Hopkins} P.~F.,  {Kere{\v{s}}} D.,  {O{\~n}orbe} J.,  {Faucher-Gigu{\`e}re}
  C.-A.,  {Quataert} E.,  {Murray} N.,   {Bullock} J.~S.,  2014, \mn@doi
  [\mnras] {10.1093/mnras/stu1738}, \href
  {https://ui.adsabs.harvard.edu/abs/2014MNRAS.445..581H} {445, 581}

\bibitem[\protect\citeauthoryear{{Hopkins} et~al.,}{{Hopkins}
  et~al.}{2018}]{2018Hopkins}
{Hopkins} P.~F.,  et~al., 2018, \mn@doi [\mnras] {10.1093/mnras/sty1690}, \href
  {https://ui.adsabs.harvard.edu/abs/2018MNRAS.480..800H} {480, 800}

\bibitem[\protect\citeauthoryear{{Hui}, {Ostriker}, {Tremaine}  \&
  {Witten}}{{Hui} et~al.}{2017}]{2017Hui}
{Hui} L.,  {Ostriker} J.~P.,  {Tremaine} S.,   {Witten} E.,  2017, \mn@doi
  [\prd] {10.1103/PhysRevD.95.043541}, \href
  {https://ui.adsabs.harvard.edu/abs/2017PhRvD..95d3541H} {95, 043541}

\bibitem[\protect\citeauthoryear{{Kamada}, {Kaplinghat}, {Pace}  \&
  {Yu}}{{Kamada} et~al.}{2017}]{2017Kamada}
{Kamada} A.,  {Kaplinghat} M.,  {Pace} A.~B.,   {Yu} H.-B.,  2017, \mn@doi
  [\prl] {10.1103/PhysRevLett.119.111102}, \href
  {https://ui.adsabs.harvard.edu/abs/2017PhRvL.119k1102K} {119, 111102}

\bibitem[\protect\citeauthoryear{{Klypin}, {Kravtsov}, {Valenzuela}  \&
  {Prada}}{{Klypin} et~al.}{1999}]{1999Klypin}
{Klypin} A.,  {Kravtsov} A.~V.,  {Valenzuela} O.,   {Prada} F.,  1999, \mn@doi
  [\apj] {10.1086/307643}, \href
  {https://ui.adsabs.harvard.edu/abs/1999ApJ...522...82K} {522, 82}

\bibitem[\protect\citeauthoryear{{Kusenko}}{{Kusenko}}{2009}]{2009Kusenko}
{Kusenko} A.,  2009, \mn@doi [\physrep] {10.1016/j.physrep.2009.07.004}, \href
  {https://ui.adsabs.harvard.edu/abs/2009PhR...481....1K} {481, 1}

\bibitem[\protect\citeauthoryear{{Lazar} et~al.,}{{Lazar}
  et~al.}{2020}]{2020Lazar}
{Lazar} A.,  et~al., 2020, \mn@doi [\mnras] {10.1093/mnras/staa2101}, \href
  {https://ui.adsabs.harvard.edu/abs/2020MNRAS.497.2393L} {497, 2393}

\bibitem[\protect\citeauthoryear{{Lovell}, {Frenk}, {Eke}, {Jenkins}, {Gao}  \&
  {Theuns}}{{Lovell} et~al.}{2014}]{2014Lovell}
{Lovell} M.~R.,  {Frenk} C.~S.,  {Eke} V.~R.,  {Jenkins} A.,  {Gao} L.,
  {Theuns} T.,  2014, \mn@doi [\mnras] {10.1093/mnras/stt2431}, \href
  {https://ui.adsabs.harvard.edu/abs/2014MNRAS.439..300L} {439, 300}

\bibitem[\protect\citeauthoryear{{Lovell} et~al.,}{{Lovell}
  et~al.}{2018}]{2018Lovell}
{Lovell} M.~R.,  et~al., 2018, \mn@doi [\mnras] {10.1093/mnras/sty2339}, \href
  {https://ui.adsabs.harvard.edu/abs/2018MNRAS.481.1950L} {481, 1950}

\bibitem[\protect\citeauthoryear{{Ludlow}, {Schaye}, {Schaller}  \&
  {Bower}}{{Ludlow} et~al.}{2020}]{2020Ludlow}
{Ludlow} A.~D.,  {Schaye} J.,  {Schaller} M.,   {Bower} R.,  2020, \mn@doi
  [\mnras] {10.1093/mnras/staa316}, \href
  {https://ui.adsabs.harvard.edu/abs/2020MNRAS.493.2926L} {493, 2926}

\bibitem[\protect\citeauthoryear{{Lynden-Bell} \& {Wood}}{{Lynden-Bell} \&
  {Wood}}{1968}]{1968Lynden-Bell}
{Lynden-Bell} D.,  {Wood} R.,  1968, \mn@doi [\mnras]
  {10.1093/mnras/138.4.495}, \href
  {https://ui.adsabs.harvard.edu/abs/1968MNRAS.138..495L} {138, 495}

\bibitem[\protect\citeauthoryear{{McGaugh}}{{McGaugh}}{2015}]{2015McGaughA}
{McGaugh} S.~S.,  2015, \mn@doi [Canadian Journal of Physics]
  {10.1139/cjp-2014-0203}, \href
  {https://ui.adsabs.harvard.edu/abs/2015CaJPh..93..250M} {93, 250}

\bibitem[\protect\citeauthoryear{{McGaugh} \& {Schombert}}{{McGaugh} \&
  {Schombert}}{2015}]{2015McGaugh}
{McGaugh} S.~S.,  {Schombert} J.~M.,  2015, \mn@doi [\apj]
  {10.1088/0004-637X/802/1/18}, \href
  {https://ui.adsabs.harvard.edu/abs/2015ApJ...802...18M} {802, 18}

\bibitem[\protect\citeauthoryear{{McGaugh}, {Schombert}, {Bothun}  \& {de
  Blok}}{{McGaugh} et~al.}{2000}]{2000McGaugh}
{McGaugh} S.~S.,  {Schombert} J.~M.,  {Bothun} G.~D.,   {de Blok} W.~J.~G.,
  2000, \mn@doi [\apjl] {10.1086/312628}, \href
  {https://ui.adsabs.harvard.edu/abs/2000ApJ...533L..99M} {533, L99}

\bibitem[\protect\citeauthoryear{{Moster}, {Naab}  \& {White}}{{Moster}
  et~al.}{2013}]{2013Moster}
{Moster} B.~P.,  {Naab} T.,   {White} S. D.~M.,  2013, \mn@doi [\mnras]
  {10.1093/mnras/sts261}, \href
  {https://ui.adsabs.harvard.edu/abs/2013MNRAS.428.3121M} {428, 3121}

\bibitem[\protect\citeauthoryear{{Muratov}, {Kere{\v{s}}},
  {Faucher-Gigu{\`e}re}, {Hopkins}, {Quataert}  \& {Murray}}{{Muratov}
  et~al.}{2015}]{2015Muratov}
{Muratov} A.~L.,  {Kere{\v{s}}} D.,  {Faucher-Gigu{\`e}re} C.-A.,  {Hopkins}
  P.~F.,  {Quataert} E.,   {Murray} N.,  2015, \mn@doi [\mnras]
  {10.1093/mnras/stv2126}, \href
  {https://ui.adsabs.harvard.edu/abs/2015MNRAS.454.2691M} {454, 2691}

\bibitem[\protect\citeauthoryear{{O{\~n}orbe}, {Boylan-Kolchin}, {Bullock},
  {Hopkins}, {Kere{\v{s}}}, {Faucher-Gigu{\`e}re}, {Quataert}  \&
  {Murray}}{{O{\~n}orbe} et~al.}{2015}]{2015Onorbe}
{O{\~n}orbe} J.,  {Boylan-Kolchin} M.,  {Bullock} J.~S.,  {Hopkins} P.~F.,
  {Kere{\v{s}}} D.,  {Faucher-Gigu{\`e}re} C.-A.,  {Quataert} E.,   {Murray}
  N.,  2015, \mn@doi [\mnras] {10.1093/mnras/stv2072}, \href
  {https://ui.adsabs.harvard.edu/abs/2015MNRAS.454.2092O} {454, 2092}

\bibitem[\protect\citeauthoryear{{Oman} et~al.,}{{Oman}
  et~al.}{2015}]{2015Oman}
{Oman} K.~A.,  et~al., 2015, \mn@doi [\mnras] {10.1093/mnras/stv1504}, \href
  {https://ui.adsabs.harvard.edu/abs/2015MNRAS.452.3650O} {452, 3650}

\bibitem[\protect\citeauthoryear{{Pillepich} et~al.,}{{Pillepich}
  et~al.}{2018a}]{2018Pillepich}
{Pillepich} A.,  et~al., 2018a, \mn@doi [\mnras] {10.1093/mnras/stx2656}, \href
  {https://ui.adsabs.harvard.edu/abs/2018MNRAS.473.4077P} {473, 4077}

\bibitem[\protect\citeauthoryear{{Pillepich} et~al.,}{{Pillepich}
  et~al.}{2018b}]{2018Pillepicha}
{Pillepich} A.,  et~al., 2018b, \mn@doi [\mnras] {10.1093/mnras/stx2656}, \href
  {https://ui.adsabs.harvard.edu/abs/2018MNRAS.473.4077P} {473, 4077}

\bibitem[\protect\citeauthoryear{{Pillepich} et~al.,}{{Pillepich}
  et~al.}{2018c}]{2018Pillepichb}
{Pillepich} A.,  et~al., 2018c, \mn@doi [\mnras] {10.1093/mnras/stx3112}, \href
  {https://ui.adsabs.harvard.edu/abs/2018MNRAS.475..648P} {475, 648}

\bibitem[\protect\citeauthoryear{{Planck Collaboration} et~al.,}{{Planck
  Collaboration} et~al.}{2014}]{2014Planck}
{Planck Collaboration} et~al., 2014, \mn@doi [\aap]
  {10.1051/0004-6361/201321591}, \href
  {https://ui.adsabs.harvard.edu/abs/2014A&A...571A..16P} {571, A16}

\bibitem[\protect\citeauthoryear{{Planck Collaboration} et~al.,}{{Planck
  Collaboration} et~al.}{2016}]{2016Planck}
{Planck Collaboration} et~al., 2016, \mn@doi [\aap]
  {10.1051/0004-6361/201526926}, \href
  {https://ui.adsabs.harvard.edu/abs/2016A&A...594A..11P} {594, A11}

\bibitem[\protect\citeauthoryear{{Polisensky} \& {Ricotti}}{{Polisensky} \&
  {Ricotti}}{2011}]{2011Polisensky}
{Polisensky} E.,  {Ricotti} M.,  2011, \mn@doi [\prd]
  {10.1103/PhysRevD.83.043506}, \href
  {https://ui.adsabs.harvard.edu/abs/2011PhRvD..83d3506P} {83, 043506}

\bibitem[\protect\citeauthoryear{{Pontzen} \& {Governato}}{{Pontzen} \&
  {Governato}}{2012}]{2012Pontzen}
{Pontzen} A.,  {Governato} F.,  2012, \mn@doi [\mnras]
  {10.1111/j.1365-2966.2012.20571.x}, \href
  {https://ui.adsabs.harvard.edu/abs/2012MNRAS.421.3464P} {421, 3464}

\bibitem[\protect\citeauthoryear{{Read}, {Agertz}  \& {Collins}}{{Read}
  et~al.}{2016}]{2016Read}
{Read} J.~I.,  {Agertz} O.,   {Collins} M.~L.~M.,  2016, \mn@doi [\mnras]
  {10.1093/mnras/stw713}, \href
  {https://ui.adsabs.harvard.edu/abs/2016MNRAS.459.2573R} {459, 2573}

\bibitem[\protect\citeauthoryear{{Robertson}, {Harvey}, {Massey}, {Eke},
  {McCarthy}, {Jauzac}, {Li}  \& {Schaye}}{{Robertson}
  et~al.}{2019}]{2019Robertson}
{Robertson} A.,  {Harvey} D.,  {Massey} R.,  {Eke} V.,  {McCarthy} I.~G.,
  {Jauzac} M.,  {Li} B.,   {Schaye} J.,  2019, \mn@doi [\mnras]
  {10.1093/mnras/stz1815}, \href
  {https://ui.adsabs.harvard.edu/abs/2019MNRAS.488.3646R} {488, 3646}

\bibitem[\protect\citeauthoryear{{Robles} et~al.,}{{Robles}
  et~al.}{2017}]{2017Robles}
{Robles} V.~H.,  et~al., 2017, \mn@doi [\mnras] {10.1093/mnras/stx2253}, \href
  {https://ui.adsabs.harvard.edu/abs/2017MNRAS.472.2945R} {472, 2945}

\bibitem[\protect\citeauthoryear{{Rocha}, {Peter}, {Bullock}, {Kaplinghat},
  {Garrison-Kimmel}, {O{\~n}orbe}  \& {Moustakas}}{{Rocha}
  et~al.}{2013}]{2013Rocha}
{Rocha} M.,  {Peter} A. H.~G.,  {Bullock} J.~S.,  {Kaplinghat} M.,
  {Garrison-Kimmel} S.,  {O{\~n}orbe} J.,   {Moustakas} L.~A.,  2013, \mn@doi
  [\mnras] {10.1093/mnras/sts514}, \href
  {https://ui.adsabs.harvard.edu/abs/2013MNRAS.430...81R} {430, 81}

\bibitem[\protect\citeauthoryear{{Sameie}, {Creasey}, {Yu}, {Sales},
  {Vogelsberger}  \& {Zavala}}{{Sameie} et~al.}{2018}]{2018Sameie}
{Sameie} O.,  {Creasey} P.,  {Yu} H.-B.,  {Sales} L.~V.,  {Vogelsberger} M.,
  {Zavala} J.,  2018, \mn@doi [\mnras] {10.1093/mnras/sty1516}, \href
  {https://ui.adsabs.harvard.edu/abs/2018MNRAS.479..359S} {479, 359}

\bibitem[\protect\citeauthoryear{{Sameie} et~al.,}{{Sameie}
  et~al.}{2021}]{2021Sameie}
{Sameie} O.,  et~al., 2021, \mn@doi [\mnras] {10.1093/mnras/stab2173}, \href
  {https://ui.adsabs.harvard.edu/abs/2021MNRAS.507..720S} {507, 720}

\bibitem[\protect\citeauthoryear{{Sawala} et~al.,}{{Sawala}
  et~al.}{2016}]{2016Sawala}
{Sawala} T.,  et~al., 2016, \mn@doi [\mnras] {10.1093/mnras/stw145}, \href
  {https://ui.adsabs.harvard.edu/abs/2016MNRAS.457.1931S} {457, 1931}

\bibitem[\protect\citeauthoryear{{Schive}, {Chiueh}  \& {Broadhurst}}{{Schive}
  et~al.}{2014}]{2014Schive}
{Schive} H.-Y.,  {Chiueh} T.,   {Broadhurst} T.,  2014, \mn@doi [Nature
  Physics] {10.1038/nphys2996}, \href
  {https://ui.adsabs.harvard.edu/abs/2014NatPh..10..496S} {10, 496}

\bibitem[\protect\citeauthoryear{{Shen}, {Hopkins}, {Necib}, {Jiang},
  {Boylan-Kolchin}  \& {Wetzel}}{{Shen} et~al.}{2021}]{2021Shen}
{Shen} X.,  {Hopkins} P.~F.,  {Necib} L.,  {Jiang} F.,  {Boylan-Kolchin} M.,
  {Wetzel} A.,  2021, \mn@doi [\mnras] {10.1093/mnras/stab2042}, \href
  {https://ui.adsabs.harvard.edu/abs/2021MNRAS.506.4421S} {506, 4421}

\bibitem[\protect\citeauthoryear{{Shen}, {Hopkins}, {Necib}, {Jiang},
  {Boylan-Kolchin}  \& {Wetzel}}{{Shen} et~al.}{2022}]{2022Shen}
{Shen} X.,  {Hopkins} P.~F.,  {Necib} L.,  {Jiang} F.,  {Boylan-Kolchin} M.,
  {Wetzel} A.,  2022, arXiv e-prints, \href
  {https://ui.adsabs.harvard.edu/abs/2022arXiv220605327S} {p. arXiv:2206.05327}

\bibitem[\protect\citeauthoryear{{Simpson}, {Grand}, {G{\'o}mez}, {Marinacci},
  {Pakmor}, {Springel}, {Campbell}  \& {Frenk}}{{Simpson}
  et~al.}{2018}]{2018Simpson}
{Simpson} C.~M.,  {Grand} R. J.~J.,  {G{\'o}mez} F.~A.,  {Marinacci} F.,
  {Pakmor} R.,  {Springel} V.,  {Campbell} D. J.~R.,   {Frenk} C.~S.,  2018,
  \mn@doi [\mnras] {10.1093/mnras/sty774}, \href
  {https://ui.adsabs.harvard.edu/abs/2018MNRAS.478..548S} {478, 548}

\bibitem[\protect\citeauthoryear{{Spergel} \& {Steinhardt}}{{Spergel} \&
  {Steinhardt}}{2000}]{2000Spergel}
{Spergel} D.~N.,  {Steinhardt} P.~J.,  2000, \mn@doi [\prl]
  {10.1103/PhysRevLett.84.3760}, \href
  {https://ui.adsabs.harvard.edu/abs/2000PhRvL..84.3760S} {84, 3760}

\bibitem[\protect\citeauthoryear{{Springel}}{{Springel}}{2010}]{2010Springel}
{Springel} V.,  2010, \mn@doi [\mnras] {10.1111/j.1365-2966.2009.15715.x},
  \href {https://ui.adsabs.harvard.edu/abs/2010MNRAS.401..791S} {401, 791}

\bibitem[\protect\citeauthoryear{{Springel} et~al.,}{{Springel}
  et~al.}{2018}]{2018Springel}
{Springel} V.,  et~al., 2018, \mn@doi [\mnras] {10.1093/mnras/stx3304}, \href
  {https://ui.adsabs.harvard.edu/abs/2018MNRAS.475..676S} {475, 676}

\bibitem[\protect\citeauthoryear{{Torrey}, {Vogelsberger}, {Genel}, {Sijacki},
  {Springel}  \& {Hernquist}}{{Torrey} et~al.}{2014}]{2014Torrey}
{Torrey} P.,  {Vogelsberger} M.,  {Genel} S.,  {Sijacki} D.,  {Springel} V.,
  {Hernquist} L.,  2014, \mn@doi [\mnras] {10.1093/mnras/stt2295}, \href
  {https://ui.adsabs.harvard.edu/abs/2014MNRAS.438.1985T} {438, 1985}

\bibitem[\protect\citeauthoryear{{Tulin} \& {Yu}}{{Tulin} \&
  {Yu}}{2018}]{2018Tulin}
{Tulin} S.,  {Yu} H.-B.,  2018, \mn@doi [\physrep]
  {10.1016/j.physrep.2017.11.004}, \href
  {https://ui.adsabs.harvard.edu/abs/2018PhR...730....1T} {730, 1}

\bibitem[\protect\citeauthoryear{{Vargya}, {Sanderson}, {Sameie},
  {Boylan-Kolchin}, {Hopkins}, {Wetzel}  \& {Graus}}{{Vargya}
  et~al.}{2021}]{2021Vargya}
{Vargya} D.,  {Sanderson} R.,  {Sameie} O.,  {Boylan-Kolchin} M.,  {Hopkins}
  P.~F.,  {Wetzel} A.,   {Graus} A.,  2021, arXiv e-prints, \href
  {https://ui.adsabs.harvard.edu/abs/2021arXiv210414069V} {p. arXiv:2104.14069}

\bibitem[\protect\citeauthoryear{{Vogelsberger}, {Zavala}  \&
  {Loeb}}{{Vogelsberger} et~al.}{2012}]{2012Vogelsberger}
{Vogelsberger} M.,  {Zavala} J.,   {Loeb} A.,  2012, \mn@doi [\mnras]
  {10.1111/j.1365-2966.2012.21182.x}, \href
  {https://ui.adsabs.harvard.edu/abs/2012MNRAS.423.3740V} {423, 3740}

\bibitem[\protect\citeauthoryear{{Vogelsberger} et~al.,}{{Vogelsberger}
  et~al.}{2014a}]{2014VogelsbergerA}
{Vogelsberger} M.,  et~al., 2014a, \mn@doi [\mnras] {10.1093/mnras/stu1536},
  \href {https://ui.adsabs.harvard.edu/abs/2014MNRAS.444.1518V} {444, 1518}

\bibitem[\protect\citeauthoryear{{Vogelsberger}, {Zavala}, {Simpson}  \&
  {Jenkins}}{{Vogelsberger} et~al.}{2014b}]{2014Vogelsberger}
{Vogelsberger} M.,  {Zavala} J.,  {Simpson} C.,   {Jenkins} A.,  2014b, \mn@doi
  [\mnras] {10.1093/mnras/stu1713}, \href
  {https://ui.adsabs.harvard.edu/abs/2014MNRAS.444.3684V} {444, 3684}

\bibitem[\protect\citeauthoryear{{Vogelsberger} et~al.,}{{Vogelsberger}
  et~al.}{2014c}]{2014Vogelsberger_nature}
{Vogelsberger} M.,  et~al., 2014c, \mn@doi [\nat] {10.1038/nature13316}, \href
  {https://ui.adsabs.harvard.edu/abs/2014Natur.509..177V} {509, 177}

\bibitem[\protect\citeauthoryear{{Vogelsberger}, {Zavala}, {Cyr-Racine},
  {Pfrommer}, {Bringmann}  \& {Sigurdson}}{{Vogelsberger}
  et~al.}{2016}]{2016Vogelsberger}
{Vogelsberger} M.,  {Zavala} J.,  {Cyr-Racine} F.-Y.,  {Pfrommer} C.,
  {Bringmann} T.,   {Sigurdson} K.,  2016, \mn@doi [\mnras]
  {10.1093/mnras/stw1076}, \href
  {https://ui.adsabs.harvard.edu/abs/2016MNRAS.460.1399V} {460, 1399}

\bibitem[\protect\citeauthoryear{{Vogelsberger}, {Marinacci}, {Torrey}  \&
  {Puchwein}}{{Vogelsberger} et~al.}{2020}]{2020Vogelsberger}
{Vogelsberger} M.,  {Marinacci} F.,  {Torrey} P.,   {Puchwein} E.,  2020,
  \mn@doi [Nature Reviews Physics] {10.1038/s42254-019-0127-2}, \href
  {https://ui.adsabs.harvard.edu/abs/2020NatRP...2...42V} {2, 42}

\bibitem[\protect\citeauthoryear{{Weinberger} et~al.,}{{Weinberger}
  et~al.}{2018}]{2018Weinberger}
{Weinberger} R.,  et~al., 2018, \mn@doi [\mnras] {10.1093/mnras/sty1733}, \href
  {https://ui.adsabs.harvard.edu/abs/2018MNRAS.479.4056W} {479, 4056}

\bibitem[\protect\citeauthoryear{{de Blok} \& {McGaugh}}{{de Blok} \&
  {McGaugh}}{1997}]{1997Blok}
{de Blok} W.~J.~G.,  {McGaugh} S.~S.,  1997, \mn@doi [\mnras]
  {10.1093/mnras/290.3.533}, \href
  {https://ui.adsabs.harvard.edu/abs/1997MNRAS.290..533D} {290, 533}

\makeatother
\end{thebibliography}




\bsp	
\label{lastpage}
\end{document}